\documentclass[10pt,twocolumn,aps,prd,preprintnumbers,showpacs,superscriptaddress,nofootinbib,amsmath,amssymb,floats,floatfix, nofootinbib,notitlepage]{revtex4-1}

\usepackage{graphicx}
\usepackage{hyperref}
\usepackage{subfigure}
\usepackage{multirow}
\usepackage[toc,page]{appendix}
\usepackage[normalem]{ulem}
\usepackage{adjustbox}
\usepackage{latexsym}
\usepackage{amsmath}
\usepackage{amssymb}
\usepackage{amsfonts}
\usepackage{times}
\usepackage{dcolumn}
\usepackage{bm}
\usepackage{tikz}
\usepackage{bigints}
\usepackage{array,tabularx,multirow,booktabs}
\usepackage{tabularx}
\usepackage{multirow}
\usepackage[tracking=true]{microtype}
\SetTracking{}{500}
\SetTracking{encoding={*}, shape=sc}{40}

\begin{document}
\title{Weak deflection angle of a dirty black hole}
\author{Reggie C. Pantig}
\email{reggie.pantig@dlsu.edu.ph}
\affiliation{Physics Department, De La Salle University-Manila, 2401 Taft Ave., 1004 Manila Philippines}
\author{Emmanuel T. Rodulfo}
\email{emmanuel.rodulfo@dlsu.edu.ph}

\begin{abstract}
In this paper, we examine the effect of dark matter to a Kerr black hole of mass $m$. The metric is derived using the Newman-Janis algorithm, where the seed metric originates from the Schwarzschild black hole surrounded by a spherical shell of dark matter with mass $M$ and thickness $\Delta r_{s}$. The seed metric is also described in terms of a piecewise mass function with three different conditions. Specializing in the non-trivial case where the observer resides inside the dark matter shell, we analyzed how the effective mass of the black hole environment affects the basic black hole properties. A high concentration of dark matter near the rotating black hole is needed to have considerable deviations on the horizons, ergosphere, and photonsphere radius. The time-like geodesic, however, shows more sensitivity to deviation even at very low dark matter density. Further, the location of energy extraction via the Penrose process is also shown to remain unchanged. With how the dark matter distribution is described in the mass function, and the complexity of how the shadow radius is defined for a Kerr black hole, deriving an analytic expression for $\Delta r_{s}$ as a condition for notable dark matter effects to occur remains inconvenient.
\end{abstract}

\maketitle

\section{Introduction}
Perhaps one of the most interesting objects in the universe is a black hole, at least for theoretical physicists, as it provides a theoretical playground in finding hints about the possible union of quantum theory and Einstein's general relativity. In this quest, a remarkable breakthrough happened in 2019, where the Event Horizon Telescope collaborative efforts successfully imaged the silhouette of the supermassive black hole at the heart of galaxy M87 using a technique called Very Long Baseline Interferometry (VLBI). Future improvements in visualizing black holes might reveal the true geometry of black holes \cite{Kumar2019,Ovgun2018,Tsukamoto2018,Lamy2018,Singh2018,Amir2016,younsi2016new,Cunha2017,Ayzenberg2018,tsupko2018notes,bambi2009apparent,johannsen2016testing,mizuno2018current,stuchlik2018light,shaikh2018shadows,Bisnovatyi-Kogan2017,Bisnovatyi-Kogan2019,Tsupko_2020,Firouzjaee2019,Bisnovatyi-Kogan2018,Amir2018,Xu2018a,Wang2017,Wang2018a,Hennigar2018,Gyulchev2018,Shaikh2018,Ohgami2016,Nedkova2013,Amir2019,Shaikh2019a,Kumar2019a,Gyulchev2019,Amarilla2012,Amarilla2013,Amarilla2015,Amarilla2018,Amarilla2010,Mars2018,Abdujabbarov2013,Abdujabbarov2017,Hou2018,Hou2018a,Haroon2019a,Cunha2016,Cunha2015,Abdujabbarov2016,Allahyari2019,Vagnozzi2019,Bambi2019,Li2020,Alexeyev2020}.

Recently, there are also efforts in analyzing black hole shadow influenced by astrophysical environments such as dark matter, dark energy, or gravitational waves \cite{Haroon2019,Konoplya2019,Wang2019,Jusufi2019,Xu2018,Wang2019a}. This coined the term "dirty black hole", introduced by Visser in 1992 \cite{PhysRevD.46.2445}. Some of these studies at least assumed some kind of configuration to the astrophysical environment, either hypothetical or coming from empirical data, that interacts with the black hole, and study its corresponding effects. These types of dirty black holes has a spacetime metric which represents a generic black hole because they didn't came from a specific field theory that satisfies the Einstein equation. Nevertheless, these metrics possess sufficient generality \cite{Nielsen2019}. There are also dirty black holes that legitimately came from the black hole solution of the Einstein equation with known interaction with the matter field. Interesting examples can be found at \cite{Shapere1991,Dowker1992,Gibbons1988,Allen1990,Galtsov1989,Lahiri1992}. Lastly, dirty black holes also do exists as solutions to non-Einstein theories such as the pseudo-complex general relativity (pcGR) \cite{PhysRevD.19.3554,Hess2009,PhysRevD.26.1858}. Interest in dirty black hole leads to the study of its quasinormal modes (also on its perturbative approach) \cite{Leung2018,PhysRevD.59.044034}, effect on gravitational waves \cite{Barausse2014,Nielsen2019}, properties of its geodesics as rotation is introduced, as well as particle collisions \cite{PhysRevD.92.044017,PhysRevD.86.084030,Zaslavskii2012,Zaslavskii2012a,Zaslavskii2015,Zaslavskii2015a,Zaslavskii2016}, regularity of its horizon \cite{PhysRevD.86.044019}, and absorption properties \cite{PhysRevD.93.024027}.

An interesting astrophysical environment is the dark matter, as it remains an elusive entity because of its non-interaction with the electromagnetic field. Its only manifestation is through the gravitational interaction with normal matter, and the strong belief of its existence can be traced to over 50 years of scientific studies about galactic rotation curves (for historical perspective, refer to Ref. \cite{Bertone2018}). Recently, gravitational lensing effects \cite{Cao2012, Vanderveld2012, Jung2019, Andrade2019, Huterer2018} are also used to probe dark matter (or dark energy) present in galaxies and clusters. 

Dark matter may also have a direct influence on the black hole geometry. In particular, distortions caused by dark matter to the shadow of a central black hole in a galaxy may serve as an alternative to Earth-based dark matter detection experiments \cite{Amare2019, Aprile2017}. It motivated a study in Ref. \cite{Konoplya2019}, which analytically estimated the specific condition for dark matter effects to occur notably in the shadow of a Schwarzschild black hole. In this paper, we use the same model for the dark matter configuration, and motivations to investigate a more realistic scenario - dark matter effects on a rotating black hole. The dark matter configuration is described only through its mass, and span that can be adjusted to determine dark matter density, hence, making it less model-dependent. While the Schwarzschild metric belongs to the vacuum solution of the Einstein equation, adding an agnostic dark matter shell model on it results to a metric that can be categorized as a generic black hole.

For the rest of this paper, Sect. \ref{sec2} introduces the Schwarzschild metric surrounded by dark matter as modeled in Ref. \cite{Konoplya2019}. In Sect. \ref{sec3}, we derive the rotating solution by using the seed metric introduced in Sect. \ref{sec2}. In Sections \ref{sec4} to \ref{sec7}, we investigate the effect of dark matter on the horizons, time-like and null circular orbits, black hole shadow, and its observables. Sect. \ref{sec8} is devoted to summarizing the results and recommendations for future studies. Lastly, we consider the +2 metric signature, and $G=c=1$.

\section{Schwarzschild black hole surrounded by dark matter} \label{sec2}
\begin{figure}
    \centering
    \includegraphics[width=\linewidth]{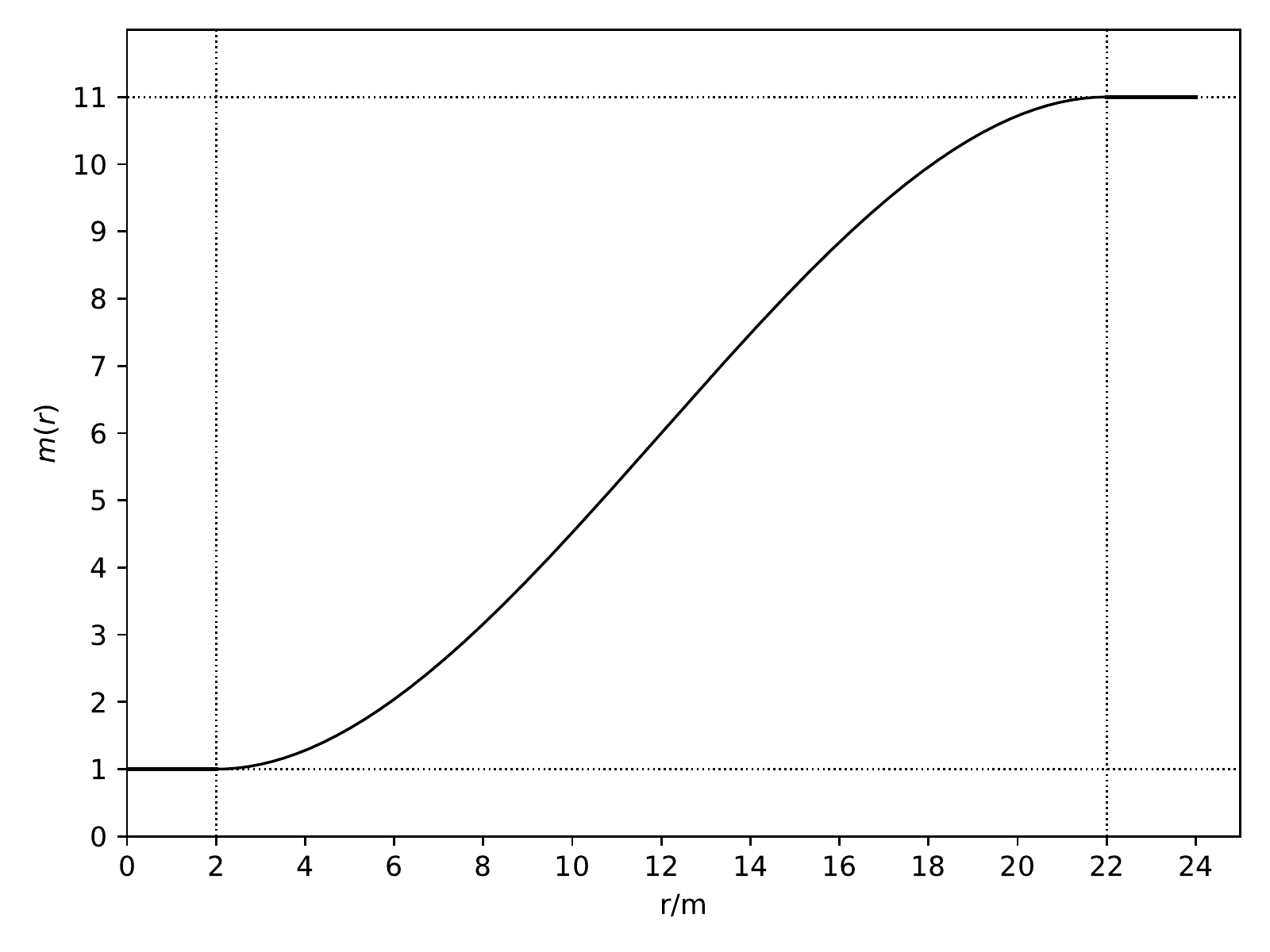}
    \caption{The choice of mass function $m(r)$. Here, $\Delta r_{s}=100m$, $r_{s}=2m$, and $M=20m$. The inner shell radius coincides with the event horizon.}
    \label{massfunc}
\end{figure}

In this section, we give a quick overview of the dirty Schwarzschild black hole presented in Ref. \cite{Konoplya2019}. The metric for a static, spherically-symmetric (SSS), uncharge, and non-rotating black hole is expressed as
\begin{equation} \label{eq1}
d{s}^2 = -f(r)dt^2 +f(r)^{-1}dr^2 +r^2\left(d\theta^2+\sin^2 \theta d\phi^2\right)
\end{equation}
where the metric function $f(r)$ is given by
\begin{equation} \label{eq2}
f(r)=1-\frac{2m}{r}.
\end{equation}
Surround the black hole of mass $m$ with a spherical shell of dark matter with inner radius $r_{s}$ and thickness $\Delta r_{s}$. The dark matter mass $M$ is also modeled as an \textit{additional} effective mass to the black hole. Moreover, the dark matter mass distribution depends on $r$-coordinate. In doing so, the metric function then becomes 
\begin{equation} \label{eq3}
f(r)=1 - \frac{2\mathcal{M}(r)}{r}
\end{equation}
where the mass function is in terms of a piecewise relation
\begin{equation} \label{eq4}
\mathcal{M}(r)=\begin{cases}
m, & r<r_{s};\\
m+ M G(r), & r_{s}\leq r\leq r_{s} +\Delta r_{s};\\
m+ M, & r>r_{s}+\Delta r_{s}.
\end{cases}
\end{equation}
and
\begin{equation} \label{eq5}
G(r)=\left(3-2\frac{r-r_{s}}{\Delta r_{s}}\right)\left(\frac{r-r_{s}}{\Delta r_{s}}\right)^{2}.
\end{equation}
In this way, when observing certain black hole phenomena, theorists can have insights or alternative perspectives where to attribute the cause of deviations from a specific theory \cite{Konoplya2019a,Cardoso2017}. Fig. \ref{massfunc} shows the plot of Eq. \eqref{eq4} for the case where the inner radius of the shell coincides with the event horizon. As shown, there is no discontinuity in $\mathcal{M}(r)$ and $\mathcal{M}'(r)$. In this paper, we will always consider the case where $r_{s}=r_{h}$ because it is more realistic compared when $r_{s}\neq r_{h}$.

\subsection{Interpretation of the domains} \label{sub2.1}
\begin{figure}
    \centering
    \includegraphics[width=\linewidth]{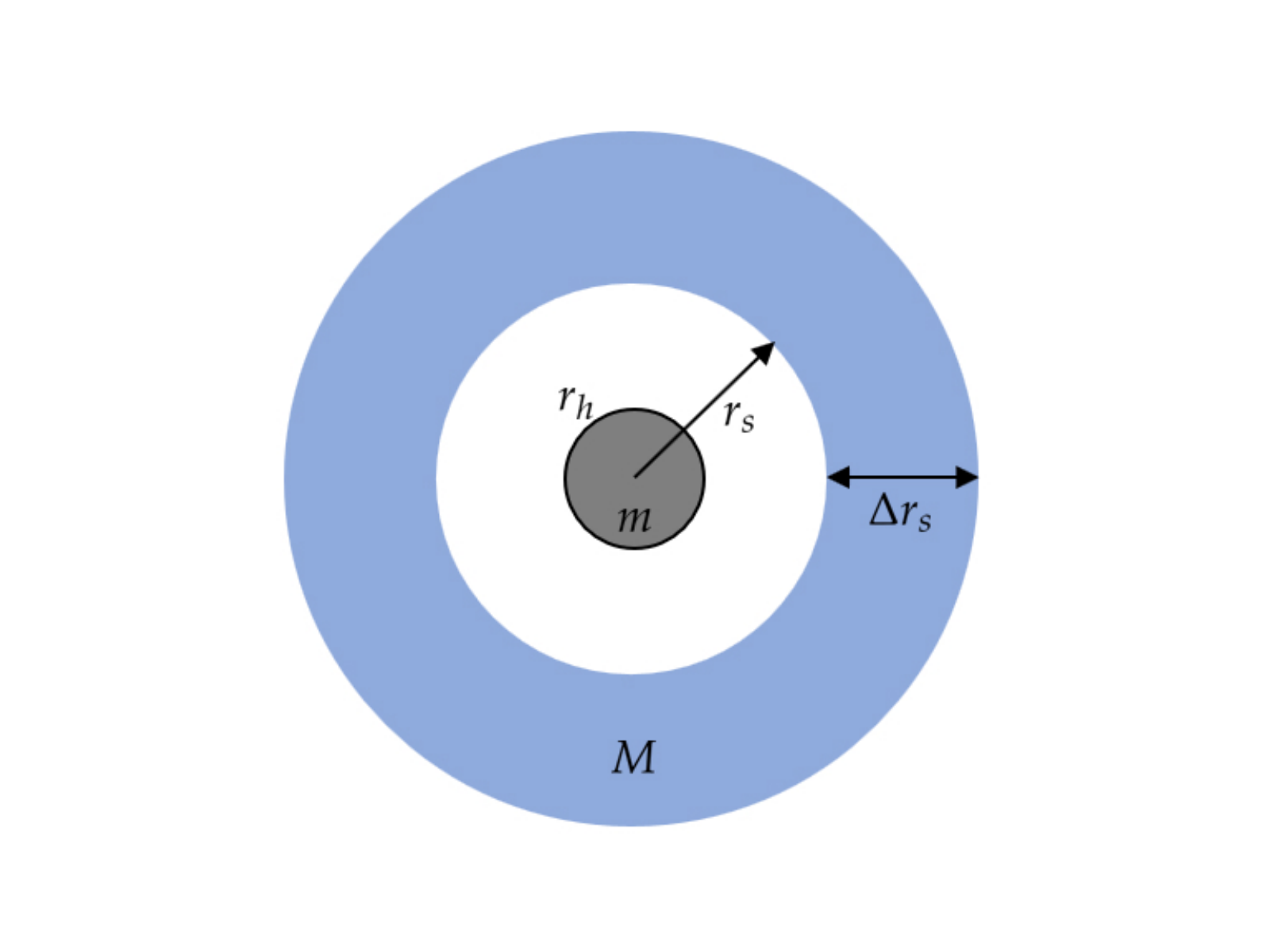}
    \caption{Schematic diagram of a black hole surrounded by a shell of dark matter where $r_{s}\neq r_{h}$. Here, if $r_{s}=10m$, $M=45m$, and $\Delta r_{s}=100m$, the horizon radius of the system is $r_{h}=3.28m$ relative to an observer inside the shell.}
    \label{bhdm}
\end{figure}
Fig. \ref{bhdm} depicts the scenario where $r_{s}\neq r_{h}$, in order to show the consequence of the first condition in Eq. \eqref{eq4}. An observer located between $r_{h}$ and $r{s}$ would then observe the pure Schwarzschild black hole, and conclude that the dark matter beyond has no effect on its geometry. The event horizon can be located at $r=2m$ and the photonsphere at $r=3m$.

An observer at $r_{s}\leq r\leq r_{s} +\Delta r_{s}$ will perceive a different location for the event horizon, as well as the photonsphere. For the horizon, it can be located by solving $r$ in the expression
\begin{equation} \label{e6}
    1 - \frac{2}{r} \left[m + M \left(3-2\frac{r-r_{s}}{\Delta r_{s}}\right)\left(\frac{r-r_{s}}{\Delta r_{s}}\right)^{2} \right]=0.
\end{equation}
In the above formula, we see that the whole system is taken into account in the calculation of the event horizon. As a result, when $r_{s} > r_{h}$, the observer inside the dark matter shell perceives a larger horizon. Such deviation seems problematic especially if the observer knows the true mass of the black hole. However, the observer just concludes that this deviation is caused by the mass of the black hole and its environment. Moreover, the deviation is not only caused solely by the mass of the black hole and dark matter below the $r$-coordinate where the observer is located, but also the dark matter mass above it. This important point was also emphasized in Ref. \cite{Konoplya2019} for the deviation for the photonsphere radius, as well as the black hole shadow.

A special case occurs when $r_{s}=r_{h}$. As one can verify, Eq. \eqref{e6} is reduced to $r=2m$, and this is regardless of where the observer is inside the dark matter shell. Here, there is no deviation in the horizon radius due to the whole system, but other phenomena such as deviation in shadow radius, and weak deflection angle can be observed \cite{Konoplya2019,Pantig2020}. This case is realistic in a sense that there's no abnormal deviation observed in the horizon radius. However, one flaw of the model is that it doesn't take into account the radial motion of the dark matter mass below the $r$-coordinate. It implies that the dark matter is static, and separate to the influence of the black hole's gravitation. Nevertheless, as emphasized again in Ref. \cite{Konoplya2019}, the utmost interest is the description of the deviation caused by the dark matter mass alone. Therefore, what is means for dark matter mass acting as an addition effective mass to the black hole is that it amplifies the effects of the black hole geometry in consideration.

As the observer is now outside the dark matter shell ($r>r_{s}+\Delta r_{s}$), the third condition in Eq. \eqref{eq4} applies. It is emphasized in Ref. \cite{Konoplya2019} that this case is an impossibility because all the dark matter mass will inevitably be absorbed by the black hole. Since dark matter is viewed as an additional effective mass, the horizon and photonsphere radius can still be computed:
\begin{equation} \label{eq6n}
    r_{h}=2(m+M), \quad \quad r_{ph}=3(m+M).
\end{equation}
It can be seen in the above expression that if the observer knows the true mass of the black hole, the deviation would be so large relative to the pure Schwarzschild case. Hence, the third condition of Eq. \eqref{eq4} is inherently unphysical, unless one analyzes the condition where all the dark matter mass is already absorbed by the black hole.

Based on the above review, it is then reasonable to take into account the second condition of Eq. \eqref{eq4} since we are interested on dark matter effects on a Kerr black hole relative to an observer inside the shell. In particular, we analyze the case where $r_{s}=r_{h}$, and $M>0$.

\section{Kerr black hole surrounded by dark matter} \label{sec3}
Using the Newman-Janis algorithm \cite{Brauer2014,canonico2011newman,Erbin2015, Erbin2017,Drake2000,Drake1997,Erbin2016,Julio2004}, we will obtain the rotating solution involving the second condition in Eq. \eqref{eq4}. The standard formalism starts with the conversion of the coordinates in Eq. \eqref{eq1} to a horizon-penetrating coordinates (also known as Eddington-Finkelstein coordinates):
\begin{equation} \label{eq7}
du=dt-dr^{*}=dt-\frac{dr}{f(r)},
\end{equation}
and we obtain
\begin{equation} \label{eq8}
d{s}^2 = -f(r)du^2 -2dudr +r^2d\theta^2 +r^2\sin^2 \theta d\phi^2.
\end{equation}
The components of the contravariant metric tensor $g^{\mu \nu}$ in line element \eqref{eq8} can be expressed in terms of the null tetrad vector components which is
\begin{equation} \label{eq9}
{g}^{\mu\nu}=-{l}^\mu{n}^\nu-{l}^\nu{n}^\mu+{m}^\mu\bar{{m}}^\nu+{m}^\nu\bar{{m}}^\mu
\end{equation}
where
\begin{align} \label{eq10}
&l=l^\mu\frac{\partial}{\partial x^\mu}=\delta^\mu_1\frac{\partial}{\partial x^\mu},\nonumber \\
&n=n^\mu\frac{\partial}{\partial x^\mu}=\left(\delta^\mu_0-\frac{f(r)}{2}\delta^\mu_1\right)\frac{\partial}{\partial x^\mu},\nonumber \\
&m=m^\mu\frac{\partial}{\partial x^\mu}=\frac{1}{\sqrt{2}r}\left(\delta^\mu_2+\frac{i}{\sin\theta}\delta^\mu_3\right)\frac{\partial}{\partial x^\mu},\nonumber \\
&\bar{m}=\bar{m}^\mu\frac{\partial}{\partial x^\mu}=\frac{1}{\sqrt{2}r}\left(\delta^\mu_2-\frac{i}{\sin\theta}\delta^\mu_3\right)\frac{\partial}{\partial x^\mu}.
\end{align}
We then do a basic complex coordinate transformation by implementing
\begin{equation} \label{eq11}
{x'}^{\mu} = x^{\mu} + ia (\delta_r^{\mu} - \delta_u^{\mu})
\cos\theta \rightarrow \\ \left\{\begin{array}{ll}
u' = u - ia\cos\theta, \\
r' = r + ia\cos\theta, \\
\theta' = \theta, \\
\phi' = \phi \end{array}\right.
\end{equation}
so that $f(r) \rightarrow f(r,\bar{r})$. Also, along with this transformation is the transformation of the null tetrad vector components via
\begin{equation} \label{eq12}
{e_a}^\mu \rightarrow {e_a^\prime}^{\mu}=\frac{{\partial{x^\prime}^\mu}}{{\partial x^\nu}}{e_a}^\nu \equiv \left({l^\prime}^\mu,{n^\prime}^\mu,{m^\prime}^\mu,{\bar{m^\prime}}^\mu\right).
\end{equation}
In particular, the transformation matrix in Eq. \eqref{eq12} is given by
\begin{equation} \label{eq13}
\left(\begin{array}{cccc}
\frac{\partial u^{\prime}}{\partial u} & \frac{\partial u^{\prime}}{\partial r} & \frac{\partial u^{\prime}}{\partial\theta} & \frac{\partial u^{\prime}}{\partial\phi}\\
\frac{\partial r^{\prime}}{\partial u} & \frac{\partial r^{\prime}}{\partial r} & \frac{\partial r^{\prime}}{\partial\theta} & \frac{\partial r^{\prime}}{\partial\phi}\\
\frac{\partial\theta^{\prime}}{\partial u} & \frac{\partial\theta^{\prime}}{\partial r} & \frac{\partial\theta^{\prime}}{\partial\theta} & \frac{\partial\theta^{\prime}}{\partial\phi}\\
\frac{\partial\phi^{\prime}}{\partial u} & \frac{\partial\phi^{\prime}}{\partial r} & \frac{\partial\phi^{\prime}}{\partial\theta} & \frac{\partial\phi^{\prime}}{\partial\phi}
\end{array}\right)=\left(\begin{array}{cccc}
1 & 0 & ia\sin\theta & 0\\
0 & 1 & -ia\sin\theta & 0\\
0 & 0 & 1 & 0\\
0 & 0 & 0 & 1
\end{array}\right)
\end{equation}
and hence, the null tetrad vector components are now the following:
\begin{align} \label{eq14}
&l^{\prime\mu}=\delta^\mu_1, \nonumber \\ 
&n^{\prime\mu}=\left(\delta^\mu_0-\frac{f(r,\bar{r})}{2}\delta^\mu_1\right),\nonumber \\
&m^{\prime\mu}=\frac{1}{\sqrt{2}\bar{r}}\left[\left(\delta^\mu_0-\delta^\mu_1\right)ia\sin\theta+\delta^\mu_2+\frac{i}{\sin\theta}\delta^\mu_3\right], \nonumber \\
&\bar{m}^{\prime\mu}=\frac{1}{\sqrt{2}r}\left[-\left(\delta^\mu_0-\delta^\mu_1\right)ia\sin\theta+\delta^\mu_2-\frac{i}{\sin\theta}\delta^\mu_3\right].
\end{align}
The components of the new contravariant metric tensor can now be constructed using
\begin{equation} \label{eq15}
{g^\prime}^{\mu\nu}=-{l^\prime}^\mu{n^\prime}^\nu-{l^\prime}^\nu{n^\prime}^\mu+{m^\prime}^\mu\bar{{m^\prime}}^\nu+{m^\prime}^\nu\bar{{m^\prime}}^\mu
\end{equation}
which results to
\begin{equation} \label{eq16}
{g^\prime}^{\mu\nu}=\left(\begin{array}{cccc}
\frac{a^{2}\sin^{2}\theta}{\Sigma} & -1-\frac{a^{2}\sin^{2}\theta}{\Sigma} & 0 & \frac{a}{\Sigma}\\
-1-\frac{a^{2}\sin^{2}\theta}{\Sigma} & F+\frac{a^{2}\sin^{2}\theta}{\Sigma} & 0 & -\frac{a}{\Sigma}\\
0 & 0 & \frac{1}{\Sigma} & 0\\
\frac{a}{\Sigma} & -\frac{a}{\Sigma} & 0 & \frac{1}{\Sigma\sin^{2}\theta}
\end{array}\right)
\end{equation}
where $\Sigma=r^2+a^2\cos^2\theta$ and $F$ is a function of $r$ and $\theta$:
\begin{equation}
    F=\frac{\Delta(r)-a^2\sin\theta^2}{r^2+a^2(1-\sin^2\theta)}.
\end{equation}
We get the inverse metric as
\begin{equation} \label{eq17}
{g^\prime}_{\mu\nu}=\left(\begin{array}{cccc}
-F & -1 & 0 & a\sin^{2}\theta\left(F-1\right)\\
-1 & 0 & 0 & a\sin^{2}\theta\\
0 & 0 & \Sigma & 0\\
a\sin^{2}\theta\left(F-1\right) & a\sin^{2}\theta & 0 & A\sin^{2}\theta
\end{array}\right)
\end{equation}
where $A=\Sigma+a^{2}\left(2-F\right)\sin^{2}\theta$. The final step in the Newman-Janis procedure is to revert back to Boyer-Lindquist coordinates by using the coordinate transformation
\begin{equation} \label{eq18}
dt=du^\prime-\frac{{r^\prime}^2+a^2}{\Delta(r)}dr^\prime, \quad d\phi=d\phi^\prime-\frac{a}{\Delta(r)}dr^\prime
\end{equation}
where $\Delta(r)$ is defined in terms of the complexified metric function $f(r,\bar{r})$. In fact, following Ref. \cite{Erbin2017} on how $r$ is complexified, we simply find that
\begin{equation} \label{eq19}
\Delta(r)=r^2-2\left[m+M\left(3-2\frac{r-r_{s}}{\Delta r_{s}}\right)\left(\frac{r-r_{s}}{\Delta r_{s}}\right)^{2}\right]r+a^2
\end{equation}
If $m(r)$ corresponds to the second condition of Eq. \eqref{eq4}, then the line element of a rotating, uncharged, and axially-symmetric black hole surrounded by a spherical shell of dark matter is given by
\begin{eqnarray} \label{eq20}
d{s}^2&=& -\left(1-\frac{2m(r)r}{\Sigma}\right)dt^2-\frac{4am(r)r \sin^2 \theta}{\Sigma}dtd\phi \nonumber \\
&&+\frac{\Sigma}{\Delta(r)}dr^2 +\Sigma d\theta^2 \nonumber \\ 
&&+\sin^2 \theta\left[r^{2}+a^{2}\left(1+\frac{2m(r)r\sin^2\theta}{\Sigma}\right)\right] d\phi^2.
\end{eqnarray}
Without a doubt, the spin parameter $a$ in the above equation involves the whole system, and its interpretation will be discussed along with the horizons in Sect. \ref{sec4}.

The original Kerr metric is a vacuum solution to the Einstein equation
\begin{equation} \label{eq22n}
    G_{\mu\nu}=8\pi T_{\mu\nu}
\end{equation}
which implies that $T_{\mu\nu}=0$, and immediately satisfies Eq. \eqref{eq22n}. If the rotating black hole is surrounded by dark matter, we can't expect that Eq. \eqref{eq20} is a vacuum solution, and dark matter's full mechanism depends on what $T_{\mu\nu}$ is. Indeed, Eq. \eqref{eq20} is a non-asymptotic spacetime which can be analyzed in a defined size of a spherical dark matter halo.

One might begin with a specific Lagrangian, solve the action, and derive the stress-energy tensor. With $T_{\mu\nu}$ at hand, one can use the Einstein equation to derive Eq. \eqref{eq20}. This is beyond the scope of this paper, however. In Ref. \cite{Azreg-Ainou2014}, it is pointed out that a rotating metric generated through NJA satisfies the Einstein equation provided that a convenient orthogonal bases are chosen. The method was also used in Ref. \cite{Jusufi2019} to show that their rotating metric, which involves the Universal Rotation Curve (URC) dark matter profile, satisfies the Einstein equation. If the Einstein equation needs to be satisfied, then \cite{Jusufi2019}
\begin{equation} \label{eq23n}
    G_{\mu\nu}-8\pi\delta_{\mu}^{\alpha}\delta_{\nu}^{\beta}G_{\alpha\beta}=0,
\end{equation}
and it follows immediately that for a non-zero stress-energy tensor,
\begin{equation} \label{eq24n}
    T_{\mu\nu}=\delta_{\mu}^{\alpha}\delta_{\nu}^{\beta}G_{\alpha\beta}.
\end{equation} 
Since the Kronecker delta is related to the orthogonal basis, we can write
\begin{equation} \label{eq25n}
    e_{a}^{\mu}e_{\alpha}^{a}T_{\mu\nu}=e_{b}^{\beta}e_{\nu}^{b}G_{\alpha\beta},
\end{equation}
and after switching indices,
\begin{equation} \label{eq26n}
    e_{a}^{\mu}e_{b}^{\nu}T_{\mu\nu}=e_{a}^{\alpha}e_{b}^{\beta}G_{\alpha\beta}.
\end{equation}
Defining $T_{ab}=e_{a}^{\mu}e_{b}^{\nu}T_{\mu\nu}$, the components of the stress-energy tensor is now given by
\begin{equation} \label{eq27n}
    T_{ab}=e_{a}^{\alpha}e_{b}^{\beta}G_{\alpha\beta}=(\rho, p_{r}, p_{\theta}, p_{\phi}).
\end{equation}
With Eq. \eqref{eq20}, the components of the Einstein tensor are given by the following:
\begin{align} \label{eq28n}
    &G_{tt}=\frac{1}{\Sigma^3}\biggl\{2m'(r)r^{2}\left[a^{2}+r(r-2m(r))\right] \nonumber \\
    &-\left[2a^{2}m'(r)\cos^{2}\theta+m''(r)\Sigma r\right]a^{2}\sin^{2}\theta\biggr\}, \nonumber \\
    &G_{t\phi}=\frac{a\sin^{2}\theta}{\Sigma^3}\biggl\{ m''(r)r\left(a^{2}+r^{2}\right)\Sigma+2m'(r)\bigl[2m(r)r^{3} \nonumber \\
    &+\left(a^{2}+r^{2}\right)a^{2}\cos^{2}\theta-r^{4}-a^{2}r^{2}\bigr]\biggl\}, \nonumber \\
    &G_{rr}=-\frac{2m'(r)r^{2}}{\Delta(r)\Sigma} \nonumber \\
    &G_{\theta \theta}=-\frac{2a^{2}m'(r)\cos^{2}\theta+a^{2}m''(r)r\cos^{2}\theta+m''(r)r^{3}}{\Sigma}, \nonumber \\
    &G_{\phi \phi}=\frac{\sin^{2}\theta}{\Sigma^{3}}\biggl\{m''(r)r\left(a^{2}+r^{2}\right)^{2}\Sigma+2a^{2}m'(r)\bigl[-a^{2}r^{2} \nonumber \\
    &+\cos^{2}\theta\left(a^{4}+3a^{2}r^{2}-2m(r)r^{3}+2r^{4}\right)+2m(r)r^{3}-r^{4}\bigr]\biggr\}.
\end{align}
For Eq. \eqref{eq27n} to apply, the convenient choice of orthogonal bases must be used:
\begin{align} \label{eq29n}
    &e_{t}^{\alpha}=\frac{1}{\sqrt{\Delta(r)\Sigma}}\left(r^{2}+a^{2},0,0,a\right), \nonumber \\
    &e_{r}^{\alpha}=\frac{\sqrt{\Delta(r)}}{\sqrt{\Sigma}}\left(0,1,0,0\right), \nonumber \\
    &e_{\theta}^{\alpha}=\frac{1}{\sqrt{\Sigma}}\left(0,0,1,0\right), \nonumber \\
    &e_{\phi}^{\alpha}=\frac{1}{\sin \theta \sqrt{\Sigma}}\left(a\sin^2 \theta,0,0,1\right).
\end{align}
With Eqs. \eqref{eq28n} and \eqref{eq29n}, the components of $T_{ab}$ can be determined:
\begin{align} \label{eq30n}
    &\rho=-p_{r}=\frac{2m'(r)r^2}{8\pi \Sigma^2}, \nonumber \\
    &p_{\theta}=p_{\phi}=p_{r}-\frac{2m'(r)+m''(r)r}{8\pi\Sigma}.
\end{align}
Indeed, the metric in Eq. \eqref{eq20} satisfies the Einstein equation. It is also clear in Eq. \eqref{eq30n} that the stress-energy tensor is anisotropic because the radial pressure is different to the tangential pressure. With the second condition in Eq. \eqref{eq4} and the parameters used in this study, one can verify that as $r\rightarrow r_{s}+\Delta r_{s}$, $p_{r}\sim 0$. Moreover, the dark matter is also pressureless for very low dark matter density (i.e. $\Delta r_{s}>>M$). For final remark, since $T_{ab}$ is not zero, it doesn't form the spacetime described by dust or perfect fluid.

\section{Horizons} \label{sec4}
We now examine the horizons of the metric given in Eq. \eqref{eq20}. Note that the metric blows up when $\Delta(r)=0$ in Eq. \eqref{eq19}, and the locations of the horizons can thus be found by solving $r$ in
\begin{equation} \label{eq21}
r^2-2\left[m+M\left(3-2\frac{r-r_{s}}{\Delta r_{s}}\right)\left(\frac{r-r_{s}}{\Delta r_{s}}\right)^{2}\right]r+a^2=0.
\end{equation}
Note that the spin parameter $a$ is for the whole system. However, even if the observer is anywhere inside the dark matter shell, what must be observed to the black hole is the spin parameter $a_{bh}$ and the event horizon $r_{h}$, which must identical to the scenario where there is no dark matter. Under the assumption such that $r_{s}=r_{h}$, this would mean that $a=a_{bh}$ and as an implication, the dark matter must not rotate with the black hole. Thus, as far as the rotation of the whole is system is concerned, it is considered as differential.

With the implementation of the Newman-Janis algorithm, the resulting model for the dirty Kerr black hole is seemingly flawed because it remained static (i.e. unaffected by its radial pull and frame-dragging). However, one must recall that originally, the seed metric came from a model treating dark matter to be separate (or unaffected) to the black hole's gravitational influence \cite{Konoplya2019}. It can be argued again that what's important is the analysis of the deviations caused by the effective mass of the dark matter. Being modeled this way, the dark matter mass can be viewed to help amplify the frame-dragging effect instead of being affected by it. Moreover, it causes changes in the black hole geometry which in turn causes deviations to the dynamics of time-like and null particles.

\begin{figure}
    \centering
    \includegraphics[width=\linewidth]{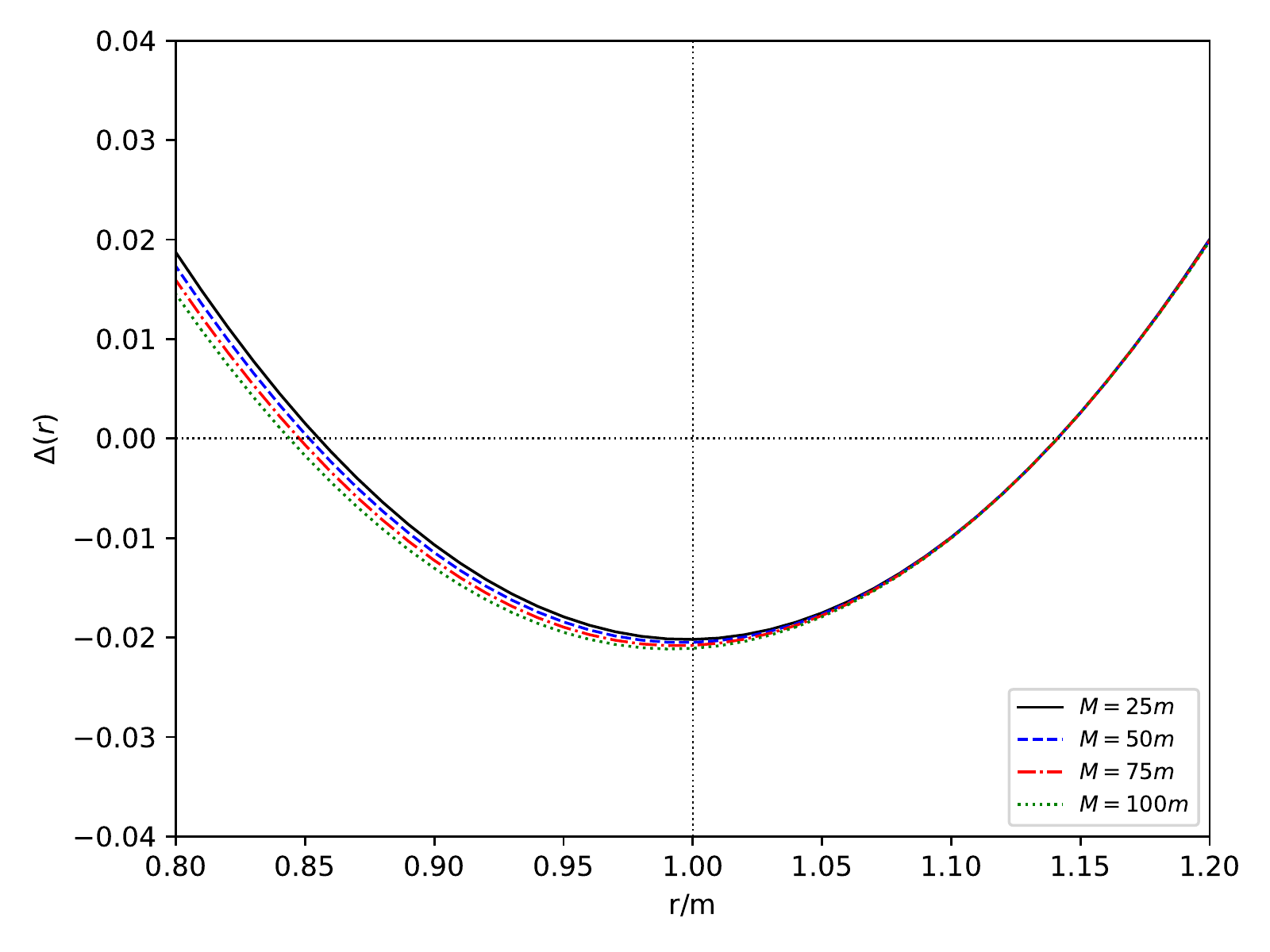}
    \caption{Behavior of $\Delta(r)=0$ for $a=0.99m$. Here, $\Delta r_{s}=100m$ and $r_{s}=1.14m$.}
    \label{fig1}
\end{figure}

Fig. \ref{fig1} shows the plot of Eq. \eqref{eq21} about the behavior of $\Delta(r)$ vs. $r$, and having interest only on how dark matter mass might affect the horizons. Note that the location of the horizons is determined when the curve intersects $\Delta(r)=0$. It can be gleaned that regardless of the dark matter mass, it has no effect on the event horizon. However, the whole system has an effect to the Cauchy horizon, which is to decrease its radius as $M$ increases. We also note that such deviation is very small for the values of $M$ considered in the plot. We also need to mention that as $M$ increases indefinitely, a third horizon at large $r$ will be produced. Such value of $M$ may serve as a constrain because such large horizon is not observable.

\begin{figure}
    \centering
    \includegraphics[width=\linewidth]{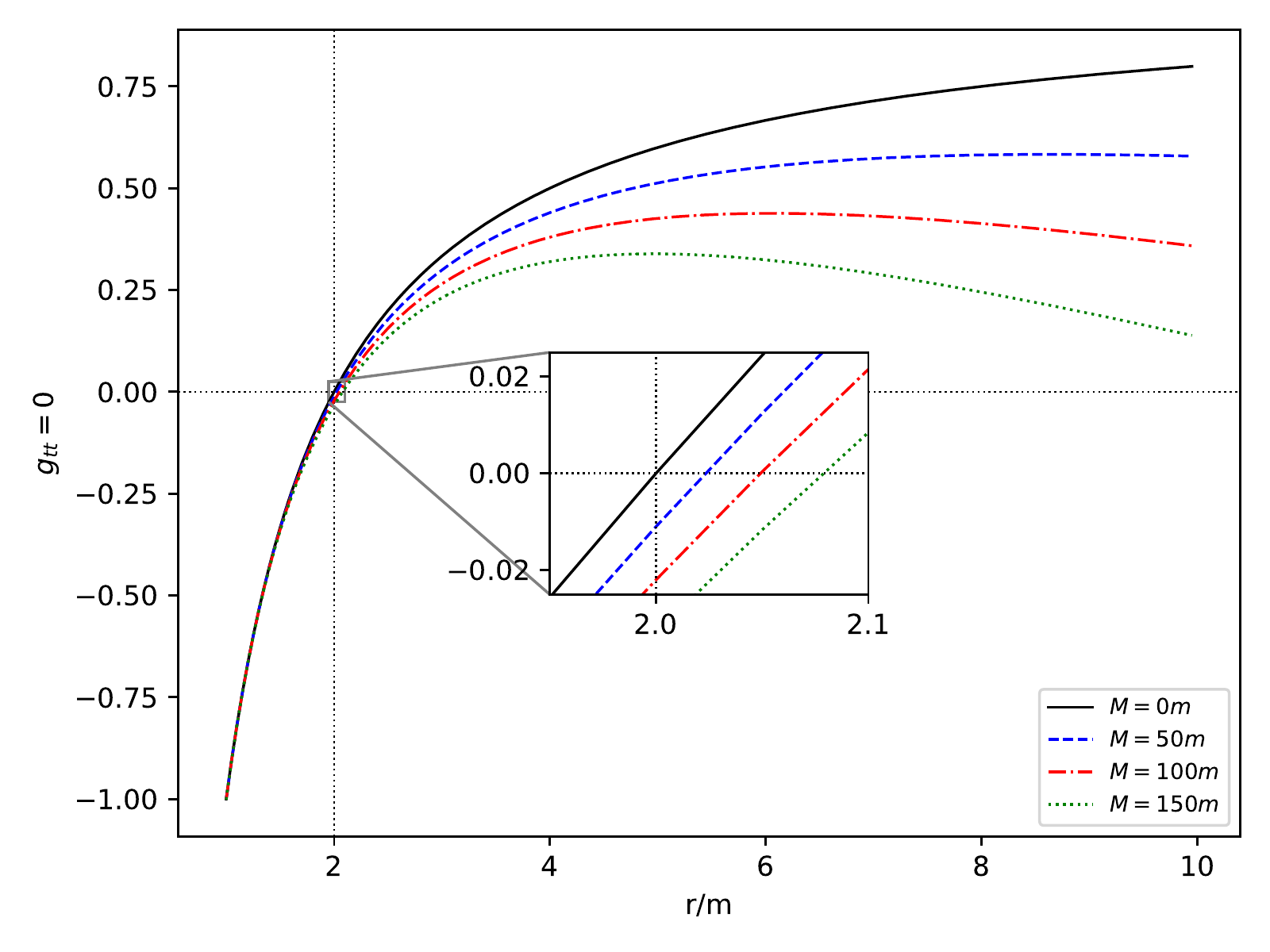}
    \caption{Behavior of $g_{tt}=0$ for $a=0.99m$. Here, $\Delta r_{s}=100m$ and $r_{s}=1.14m$.}
    \label{fig2}
\end{figure}

For the radius of the ergosphere, one can find its location by solving $r$ when $g_{tt}=0$. In particular,
\begin{equation}\label{eq-ergo}
    1-\frac{2r}{\Sigma}\left[m + M\left(3-2\frac{r-r_{s}}{\Delta r_{s}}\right)\left(\frac{r-r_{s}}{\Delta r_{s}}\right)^{2}\right]=0.
\end{equation}
Without loss of generality, if $\theta=\pi/2$, the Kerr black hole without dark matter surrounding it will give only one value of the ergoregion, which is at $r=2m$ because $g_{tt}$ becomes independent of $a$. When there is dark matter, the influence of the spin parameter $a$ remains because of $r_{s}$ depends on it. Hence, we observe that increasing the dark matter mass also increases the radius of the ergosphere as shown in Fig. \ref{fig2}. Like the horizons, extending the plot for large values of $r$ shows that $M$ is constrained because of the possible existence of a second ergoregion. Moreover, we note that deviation is very small even for high dark matter mass.

\section{Time-like circular orbits} \label{sec5}
The geodesics of both particles and photons can be studied using the Hamilton-Jacobi approach to general relativity. The Hamilton-Jacobi equation reads
\begin{equation} \label{eq22}
\frac{\partial S}{\partial \lambda}=-H
\end{equation}
where $S$ is the Jacobi action and defined in terms of an affine parameter $\lambda$ and coordinates $x^{\mu }$. In general relativity, the Hamiltonian is given by
\begin{equation} \label{eq23}
H=\frac{1}{2}g^{\mu \nu }\frac{%
\partial S}{\partial x^{\mu }}\frac{\partial S}{\partial x^{\nu }}
\end{equation}
and it follows that
\begin{equation} \label{eq24}
\frac{\partial S}{\partial \lambda }=-\frac{1}{2}g^{\mu \nu }\frac{
\partial S}{\partial x^{\mu }}\frac{\partial S}{\partial x^{\nu }}.
\end{equation}

The metric in Eq. \eqref{eq20} is independent on $t$, $\phi$, and $\lambda$, thus we can use the separability anzats given by
\begin{equation} \label{eq25}
S=\frac{1}{2}\mu ^{2}\lambda -Et+L\phi +S_{r}(r)+S_{\theta }(\theta),
\end{equation}
where $\mu $ is proportional to the particle's rest mass and $S_{r}(r)+S_{\theta }(\theta )$ are both functions of $r$ and $\theta $. The equations of motions are then derived by combining Eqs. \eqref{eq24} and \eqref{eq25}. The results are
\begin{align} \label{eq26}
&\Sigma\frac{dt}{d\lambda}=\frac{r^2+a^2}{\Delta(r)}P(r)-a(aE\sin^2\theta-L), \nonumber \\
&\Sigma\frac{dr}{d\lambda}=\sqrt{R(r)}, \nonumber \\
&\Sigma\frac{d\theta}{d\lambda}=\sqrt{\Theta(\theta)}, \nonumber \\
&\Sigma\frac{d\phi}{d\lambda}=\frac{a}{\Delta(r)}P(r)-\left(aE-\frac{L}{\sin^2\theta}\right),
\end{align}
with $P(r)$, $R(r)$ and $\Theta(\theta)$ given by
\begin{align} \label{eq27}
&P(r)=E(r^2+a^2)-aL,\nonumber \\
&R(r)=P(r)^2-\Delta(r)[Q+(aE-L)^2+\mu ^2r^2], \nonumber \\
&\Theta(\theta)=Q-\left[a^{2}\left(\mu^{2}-E^{2}\right)+\frac{L^{2}}{\sin^{2}\theta}\right] \cos^2\theta,
\end{align}
with $Q$ being the Carter constant: $Q\equiv K-(L-aE)^{2}$ and $K$ is another constant of motion.

For circular orbits, the conditions
\begin{equation} \label{eq39n}
R(r)=\frac{dR(r)}{dr}\mid_{r=r_{o}}=0
\end{equation}
must be satisfied. In order to derive the energy required for a particle to undergo circular motion, we rewrite $R(r)$ in Eq. \eqref{eq27} as we note that $X=L-aE$ \cite{Slany2013}, and $\mu=1$:
\begin{equation} \label{eq40n}
    R(r)=-2 a E  r^2 X+X^2 \left(a^2-\Delta (r)\right)+E ^2 r^4-r^2 \Delta (r)=0,
\end{equation}
while its derivative with respect to $r$ is given by
\begin{align} \label{eq41n}
    R'(r)&=-4 a E  r X+4 E ^2 r^3-r^2 \Delta '(r) \nonumber \\
    &-2 r \Delta (r)-X^2 \Delta '(r)=0
\end{align}
Eliminating the first term, we find an expression for $X^2$ \cite{raine2014black}:
\begin{equation} \label{eq42n}
    X^2=\frac{r^3 \left(\Delta '(r)-2 E ^2 r\right)}{-2 a^2-r \Delta '(r)+2 \Delta (r)}.
\end{equation}
Substituting Eq. \eqref{eq42n} to Eq. \eqref{eq41n}, and solving for $E^2$, we find the energy necessary for circular motion:
\begin{align} \label{eq43n}
    &E_{cir}^2=\frac{1}{A r^2}\biggl\{8 a^4 \Delta (r)-a^2 \bigl(r^2 \Delta '(r)^2-2 r \Delta (r) \Delta '(r)\nonumber \\
    &+16 \Delta (r)^2\bigr)\pm2 \sqrt{2} a \Delta (r) \bigl(2 a^2+r \Delta '(r) \nonumber \\
    &-2 \Delta (r)\bigr)^{3/2}-2 r \Delta (r)^2 \Delta '(r)+8 \Delta (r)^3\biggr\}
\end{align}
where $A=\left(r \Delta '(r)-4 \Delta (r)\right)^2-16 a^2 \Delta (r)$. It turns out that Eq. \eqref{eq43n} represents four equations for energy: two different values of particle's positive energies, and two different values of negative energies. In the derivation for the particle's innermost stable circular orbit (ISCO), we use the positive solutions.

The stability of the circular orbit is imprinted in $R''(r)=0$. Since Eq. \eqref{eq42n} contains the information of $R'(r)$, we can differentiate it again and obtain the expression
\begin{align} \label{eq44n}
    &r \left(\Delta (r)-a^2\right) \left(\Delta ''(r)-8 E ^2\right)+3 \left(\Delta (r)-a^2\right) \Delta '(r) \nonumber \\
    &-E ^2 r^3 \Delta ''(r)+5 E ^2 r^2 \Delta '(r)-2 r \Delta '(r)^2=0.
\end{align}
Isolating $E^2$, we obtain the energy of a particle located at the ISCO radius:
\begin{align} \label{eq45n}
    &E_{isco}^2=\frac{1}{B r}\biggl\{a^2 \left[-\left(r \Delta ''(r)+3 \Delta '(r)\right)\right]+r \Delta (r) \Delta ''(r) \nonumber \\
    &-2 r \Delta '(r)^2+3 \Delta (r) \Delta '(r)\biggr\}
\end{align}
where $B=-8 a^2+r \left(r \Delta ''(r)-5 \Delta '(r)\right)+8 \Delta (r)$. The ISCO radius can be found by equating Eq. \eqref{eq43n} and Eq. \eqref{eq45n}. We then find
\begin{align} \label{eq46n}
    &\pm2 \Delta (r) \left(a^2-\Delta (r)\right)^2\pm\frac{9}{4} r \Delta (r) \left(a^2-\Delta (r)\right) \Delta '(r) \nonumber \\
    &\pm\frac{1}{16} r^3 \Delta '(r) \left(\Delta (r) \Delta ''(r)-2 \Delta '(r)^2\right)\nonumber \\
    &\pm\frac{1}{16} r^2 \bigl[4 \Delta (r) \left(a^2-\Delta (r)\right) \Delta ''(r)+\left(15 \Delta (r)-4 a^2\right) \Delta '(r)^2\bigr] \nonumber \\
    &+a \Delta (r) \sqrt{4 a^2+2 r \Delta '(r)-4 \Delta (r)} \bigl[-a^2+\frac{1}{8} r \bigl(r \Delta ''(r)\nonumber \\
    &-5 \Delta '(r)\bigr)+\Delta (r)\bigr]=0.
\end{align}
If we use the original Kerr metric, where $\Delta(r)=r^2-2mr+a^2$, Eq. \eqref{eq46n} reduces to
\begin{equation} \label{eq47n}
    \left[3 a^2\mp8 a \sqrt{m} \sqrt{r}+r (6 m-r)\right]=0.
\end{equation}
The upper sign in Eq. \eqref{eq46n}, which resulted from the lower sign of Eq. \eqref{eq43n}, gives the upper sign in Eq. \eqref{eq47n}. The solution for $r$ when $a=m$ gives the prograde circular orbit for time-like particles ($r=m$). When the lower sign is used, the result is the retrograde orbit of the time-like particle ($r=9m$).
\begin{figure}
    \begin{minipage}{\columnwidth}
    \centering
    \includegraphics[width=\linewidth]{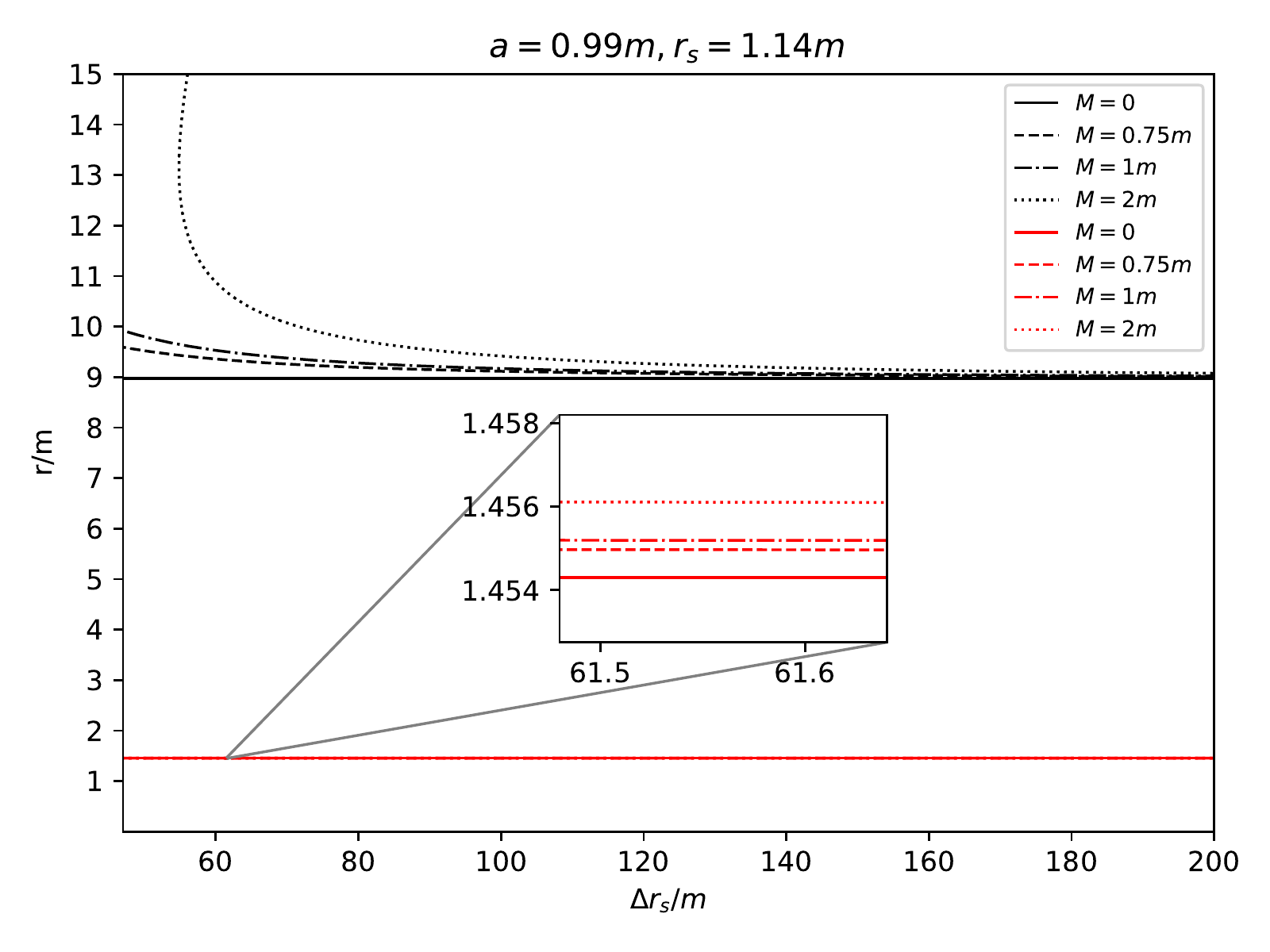}
    \end{minipage}
    \begin{minipage}{\columnwidth}
    \centering
    \includegraphics[width=\linewidth]{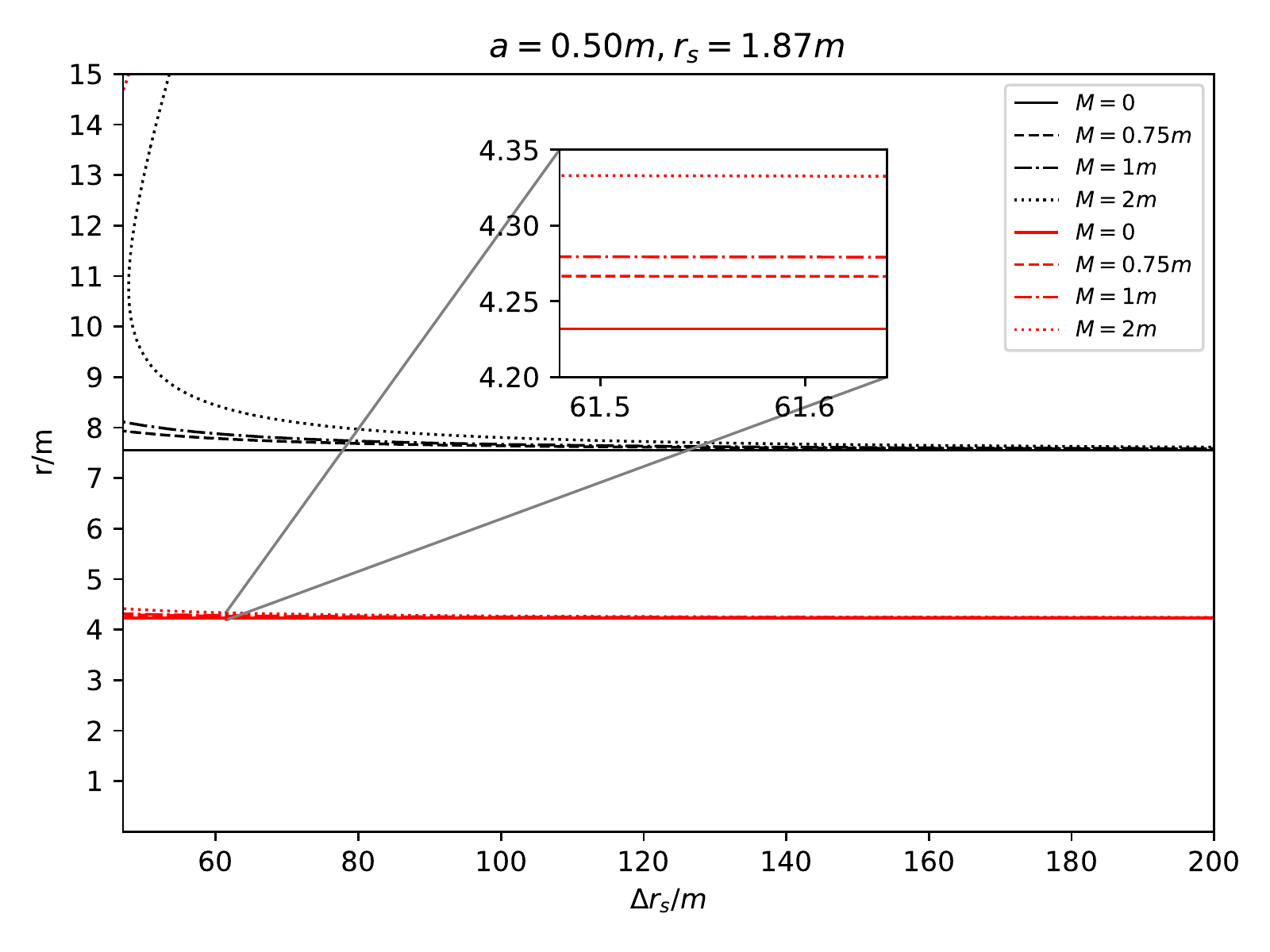}
    \end{minipage}
    \caption{Location of innermost stable circular orbit (ISCO) for a time-like particle.}
    \label{fig3}
\end{figure}
\begin{figure}
    \begin{minipage}{\columnwidth}
    \centering
    \includegraphics[width=\linewidth]{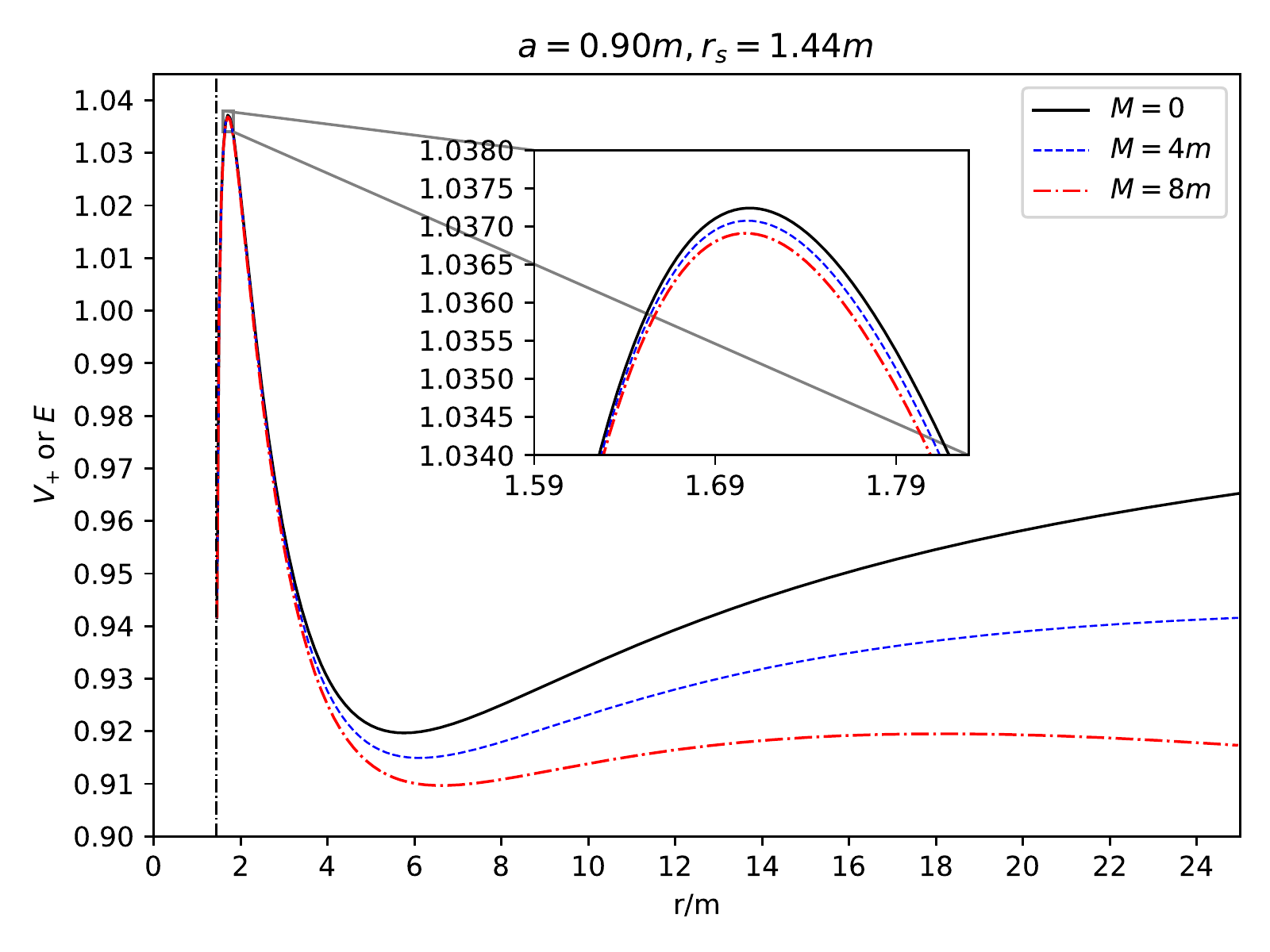}
    \end{minipage}
    \begin{minipage}{\columnwidth}
    \centering
    \includegraphics[width=\linewidth]{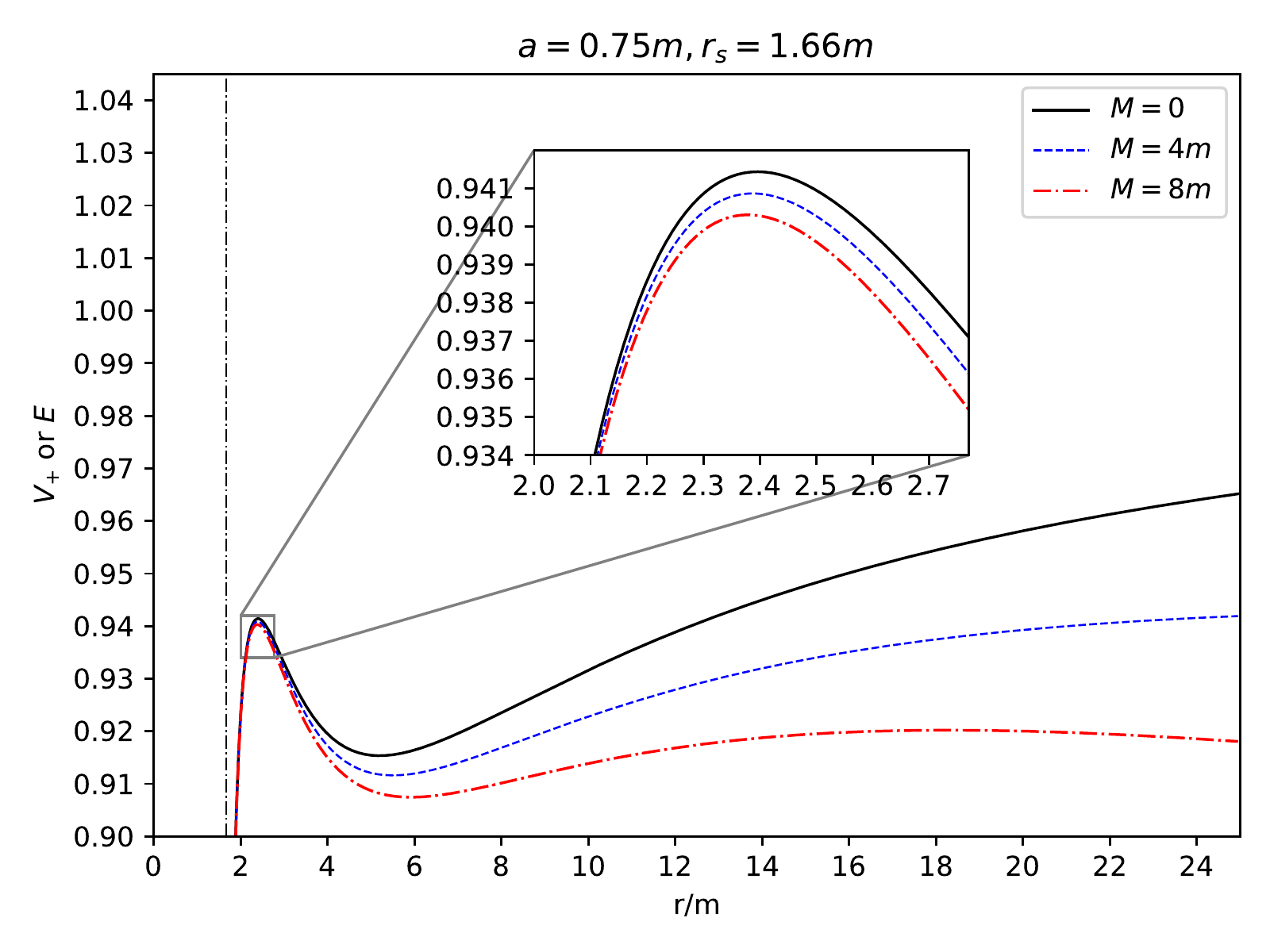}
    \end{minipage}
    \caption{Effective potential for $\Delta r_{s}=100m$ and $L=2.75m$.}
    \label{fig4}
\end{figure}

Using the metric in Eq. \eqref{eq20}, the expression that contains the information about the locations of prograde and retrograde orbits proves to be formidable. It is unfortunate that an analytic solution is inconvenient to be displayed given that $\Delta(r)$ depends on $G(r)$ in Eq. \eqref{eq4}, which gives the additional complexity. Nevertheless, numerical analysis can be implemented.

In Fig. \ref{fig3}, we plot Eq. \eqref{eq46n} in order to show the location of time-like orbits for two specific values of $a$. The inset plot shows the prograde orbit. In the near extreme case, the retrograde orbit shows more sensitivity to change than the prograde orbit. At low black hole spin, the reverse happens. Regardless of black hole spin, the retrograde orbit shows more change in its radius than the prograde orbit. It is clear that in the presence of dark matter, even at very low density, it can have some noticeable effect in the time-like orbits. With the chosen value of $M$, the retrograde orbit's radius attains an abnormality in its value, which indicates another constrain to dark matter density. Finally, these orbits are asymptotic to their corresponding values in the original Kerr black hole, and when $a=0$, the two orbits coincide at $r=6m$.

Other types of circular orbits such as bound, stable, and unstable circular orbits can be studied qualitatively using the effective potential method. Following Ref. \cite{Bautista-Olvera2019}, the effective potential in terms of angular momentum per unit mass is given by
\begin{multline} \label{eq34}
V_{\pm}=\frac{2m(r)aL}{r^{3}+a^{2}\left(2m(r)+r\right)}\\
\pm\left\{ \frac{\Delta(r)\left[(r^{2}+a^{2})^{2}-a^{2}\Delta(r)+r^{2}L^{2}\right]}{\left[r^{3}+a^{2}\left(2m(r)+r\right)\right]^{2}}\right\} ^{1/2}.
\end{multline}

Fig. \ref{fig4} tells us plenty of information about other types of circular orbits. Here, the vertical line represents the location of the event horizon. The maxima of the effective potential represents the unstable circular orbit in which any perturbation in the particle's orbit will dictate whether it will fall into the black hole or escape to infinity. The higher the spin of the black hole, the higher the energy requirement in this unstable orbit. The inset plot reveals that increasing dark matter mass decreases slightly the energy required in such an orbit. Furthermore, the radius where the peak is located decreases. The reverse happens in the stable circular orbit in which the radius where the minima occurs increases. Also, like the unstable orbit, the energy requirement decreases more obviously. For the low spin parameter, the available energy for elliptic bound orbits to occur is minimal, hence its easier to plunge into the black hole.

\begin{figure}
    \begin{minipage}{\columnwidth}
    \centering
    \includegraphics[width=\linewidth]{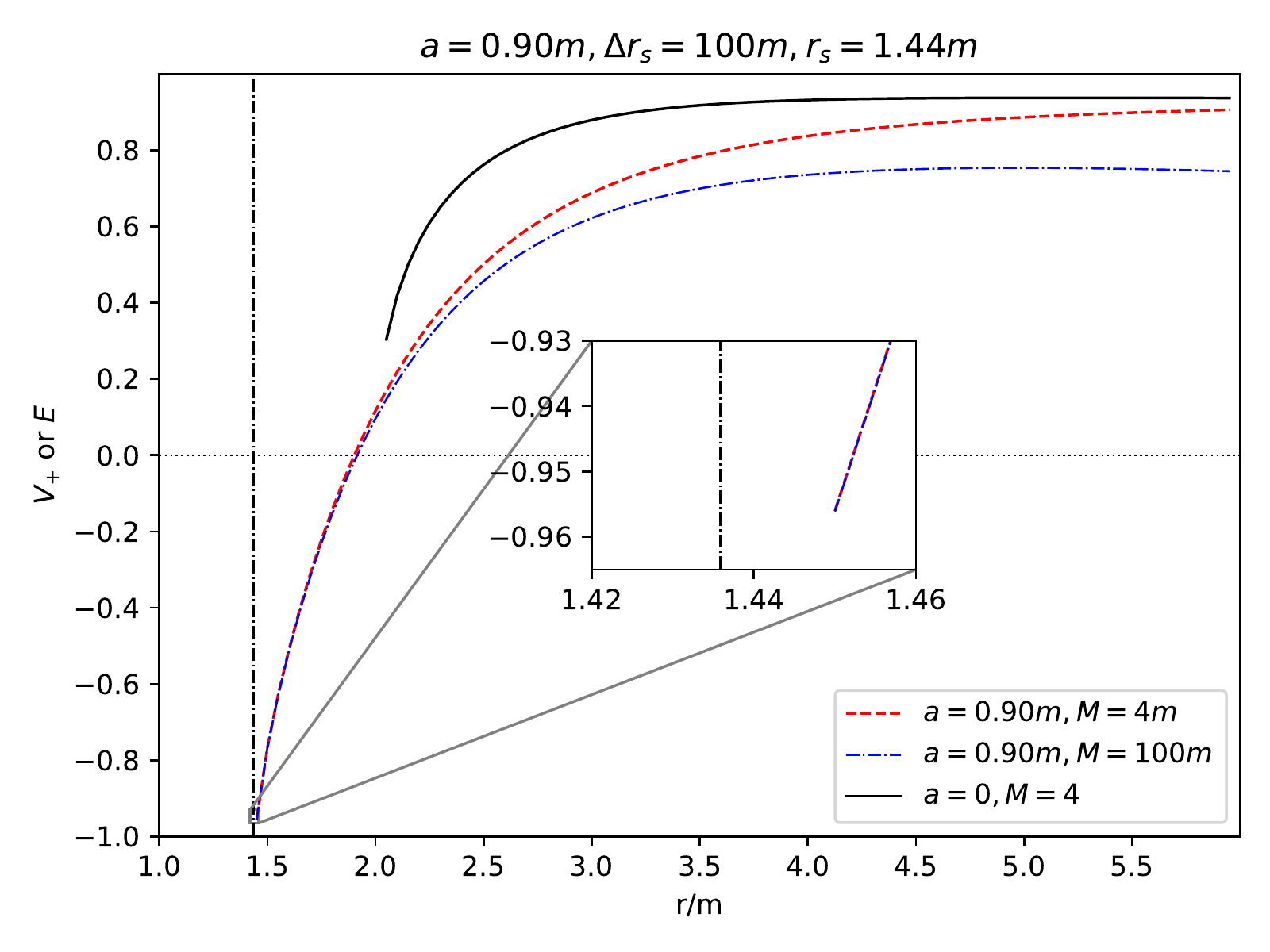}
    \end{minipage}
    \begin{minipage}{\columnwidth}
    \centering
    \includegraphics[width=\linewidth]{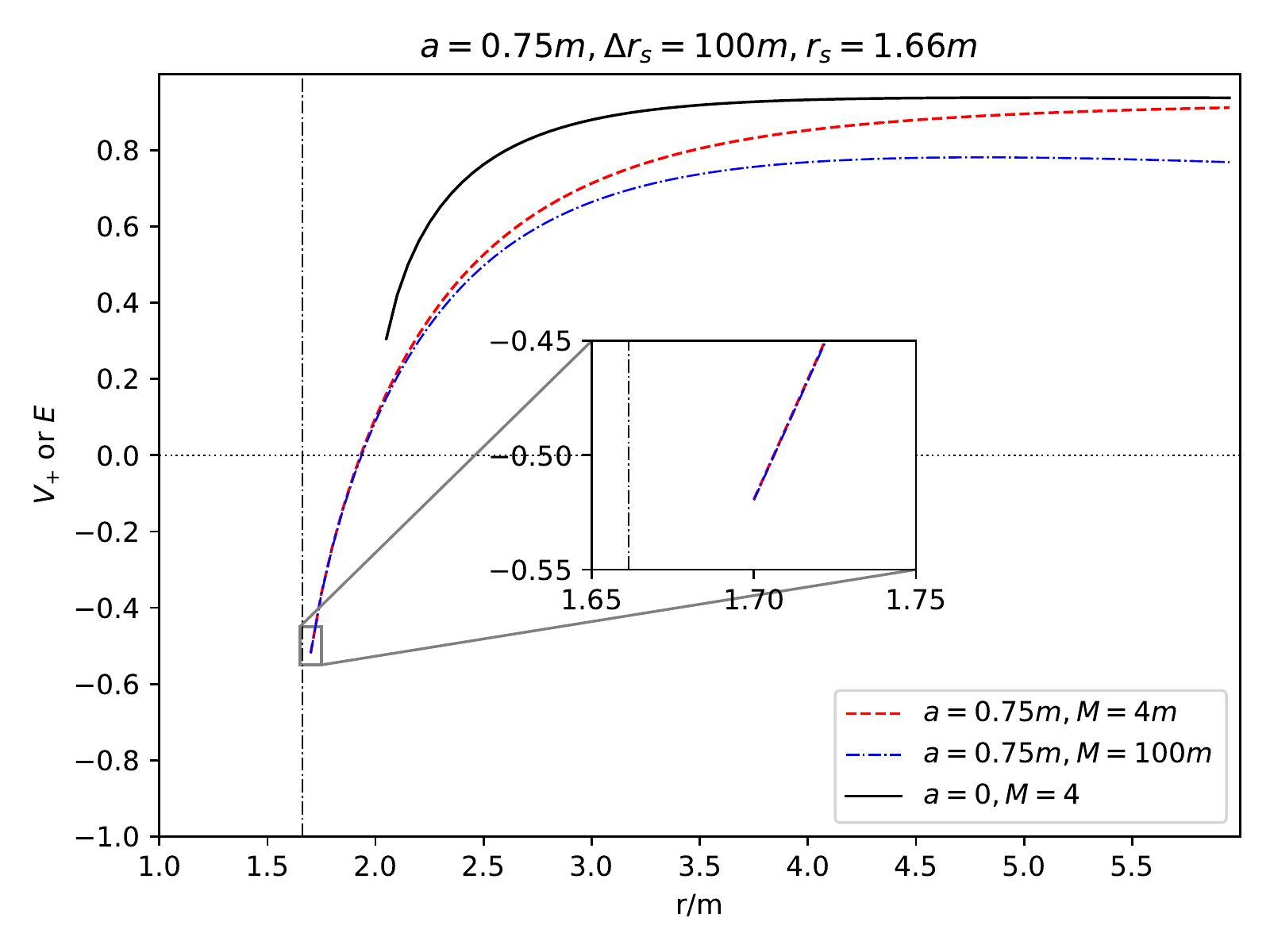}
    \end{minipage}
    \caption{Effective potential for $-L$.}
    \label{fig5}
\end{figure}

By convention, particles that revolve clockwise on a black hole have negative angular momentum $L$. Fig. \ref{fig5} shows that in a Schwarzschild case, the effective potential is always positive, and negative energy is not even allowed. If the black hole is rotating, negative effective potentials (or negative energies) of particles are allowed for $aL<0$ and energy extraction from the black hole is allowed via the Penrose process. Regardless of dark matter density, the location where the Penrose process should occur remains unchanged. For a black hole that has a very high spin (top), the particle has more negative energy compared to a black hole that spins slowly (bottom). Hence, any deviations in the Penrose process due to dark matter is very negligible.

\section{Null circular orbits and black hole shadow} \label{sec6}
\begin{figure}
    \begin{minipage}{\columnwidth}
    \centering
    \includegraphics[width=\linewidth]{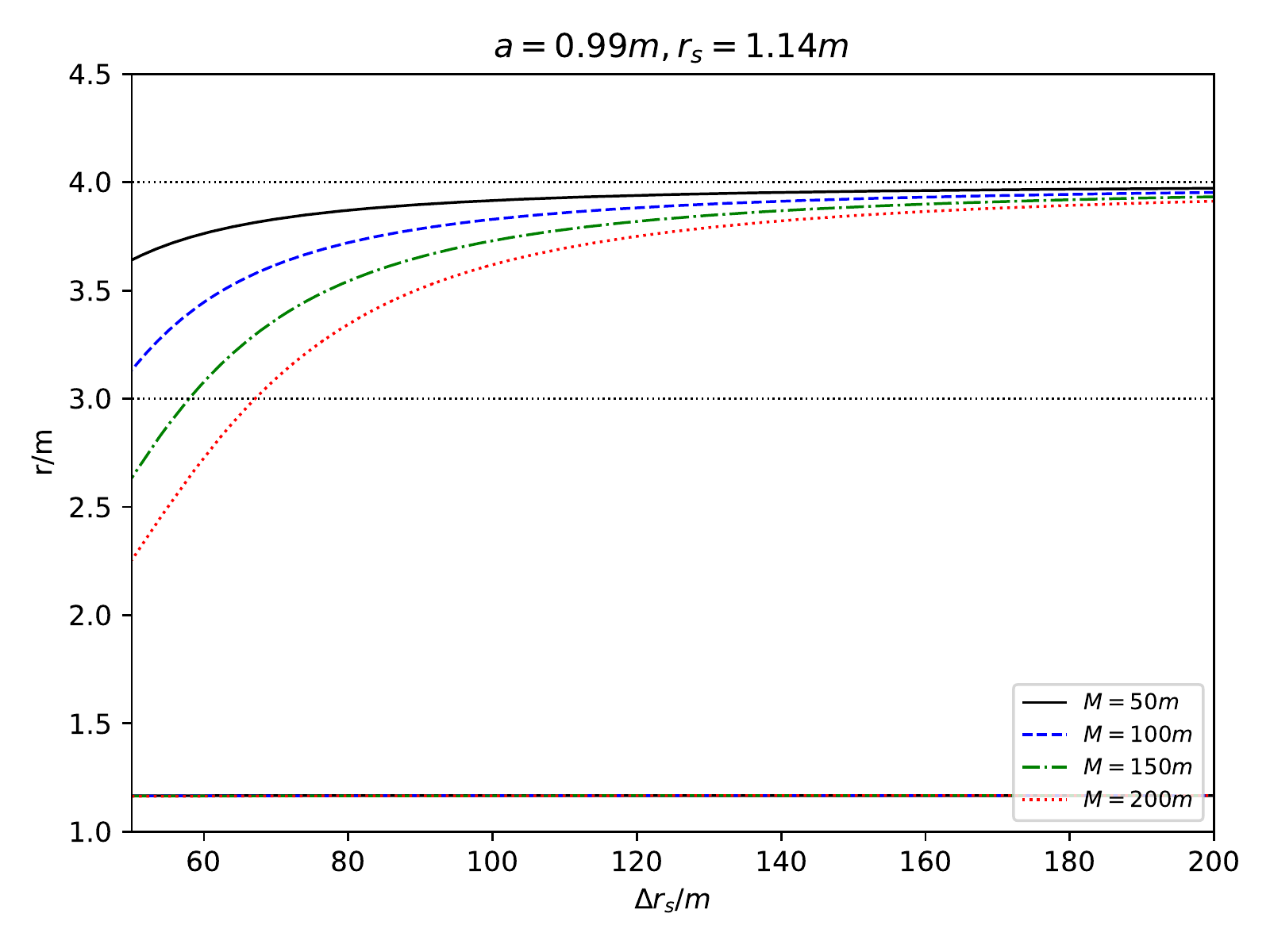}
    \end{minipage}
    \begin{minipage}{\columnwidth}
    \centering
    \includegraphics[width=\linewidth]{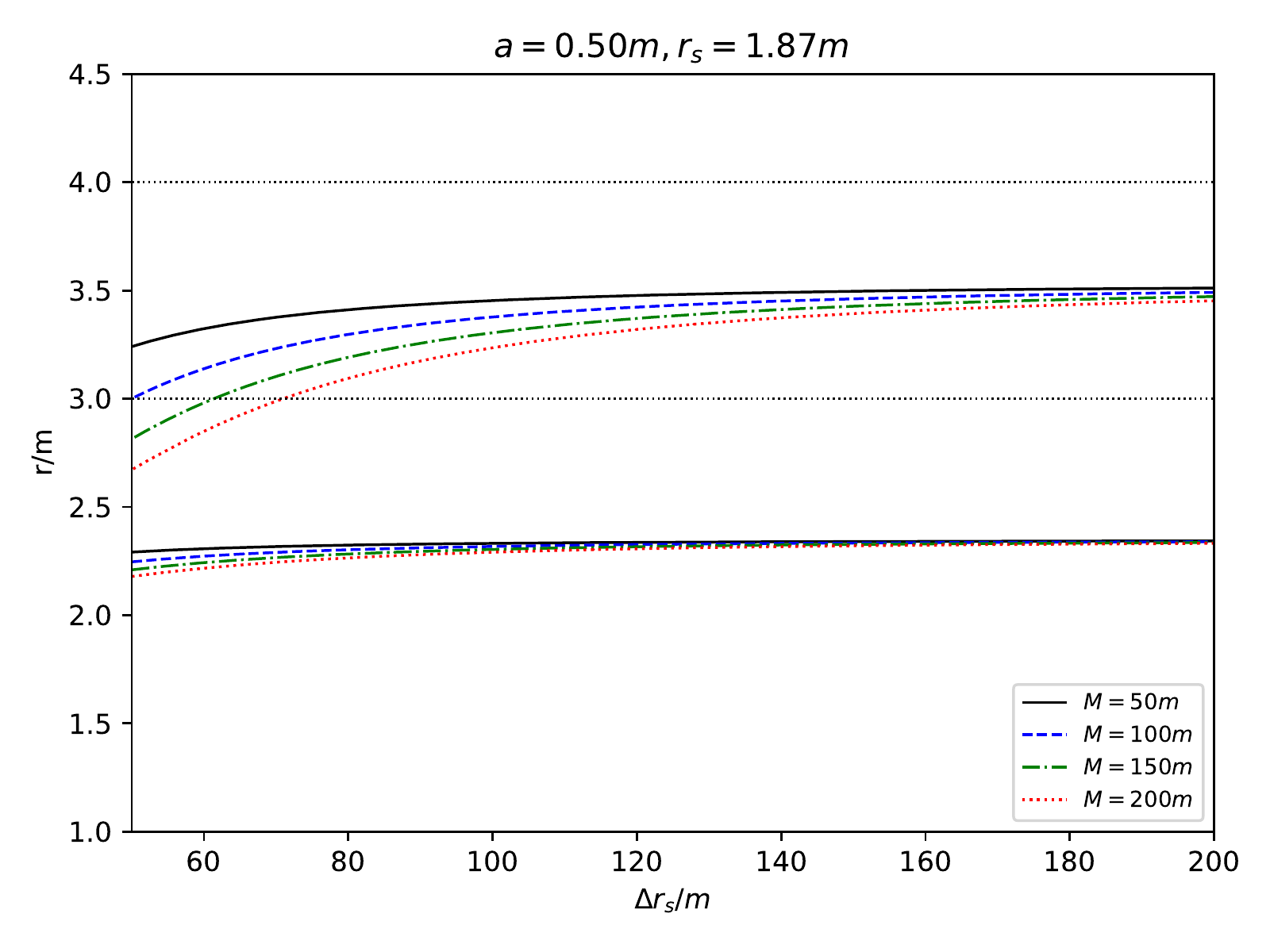}
    \end{minipage}
    \caption{Location of unstable photon orbit (Photonsphere).}
    \label{fig6}
\end{figure}
Null geodesics are of importance in studying the contour of a black hole silhouette. In particular, we need to determine and locate the unstable circular orbit for photons and do the backward ray tracing method to plot the contour of the resulting shadow in the celestial coordinates of a remote observer. For this case, we use the following impact parameters \cite{taylor2000exploring,bardeen1973timelike,teo2003spherical} and set $\mu=0$:
\begin{equation} \label{eq49n}
    \xi=\frac{L}{E}, \quad \eta=\frac{Q}{E^2}.
\end{equation}
Inserting these to $R(r)$ in Eq. \eqref{eq27}, and using the condition in Eq. \eqref{eq39n}, we obtain the following:
\begin{equation} \label{eq35}
\xi=\frac{\Delta'(r)(r^{2}+a^{2})-4\Delta(r)r}{a\Delta'(r)},
\end{equation}
\begin{equation} \label{eq36}
\eta=\frac{-r^{4}\Delta'(r)^{2}+8r^{3}\Delta(r)\Delta'(r)+16r^{2}\Delta(r)(a^{2}-\Delta(r))}{a^{2}\Delta'(r)^{2}}
\end{equation}
which appears to be general. If $M=0$, one can obtain the very well known analytic formula for the prograde and retrograde orbit radii. However, it can be tedious or inconvenient to obtain an analytic formula for a Kerr black hole with dark matter configuration given in Eq. \eqref{eq4} since the result of $\eta=0$ involves a 5th power polynomial:
\begin{equation} \label{eqr^5}
    16\Delta(r)r^{2}\left(a^{2}-\Delta(r)\right)-\Delta^\prime(r)^{2}r^{4}+8\Delta^\prime(r)\Delta(r)r^{3}=0.
\end{equation}
\begin{figure}
    \begin{minipage}{\columnwidth}
    \centering
    \includegraphics[width=\linewidth]{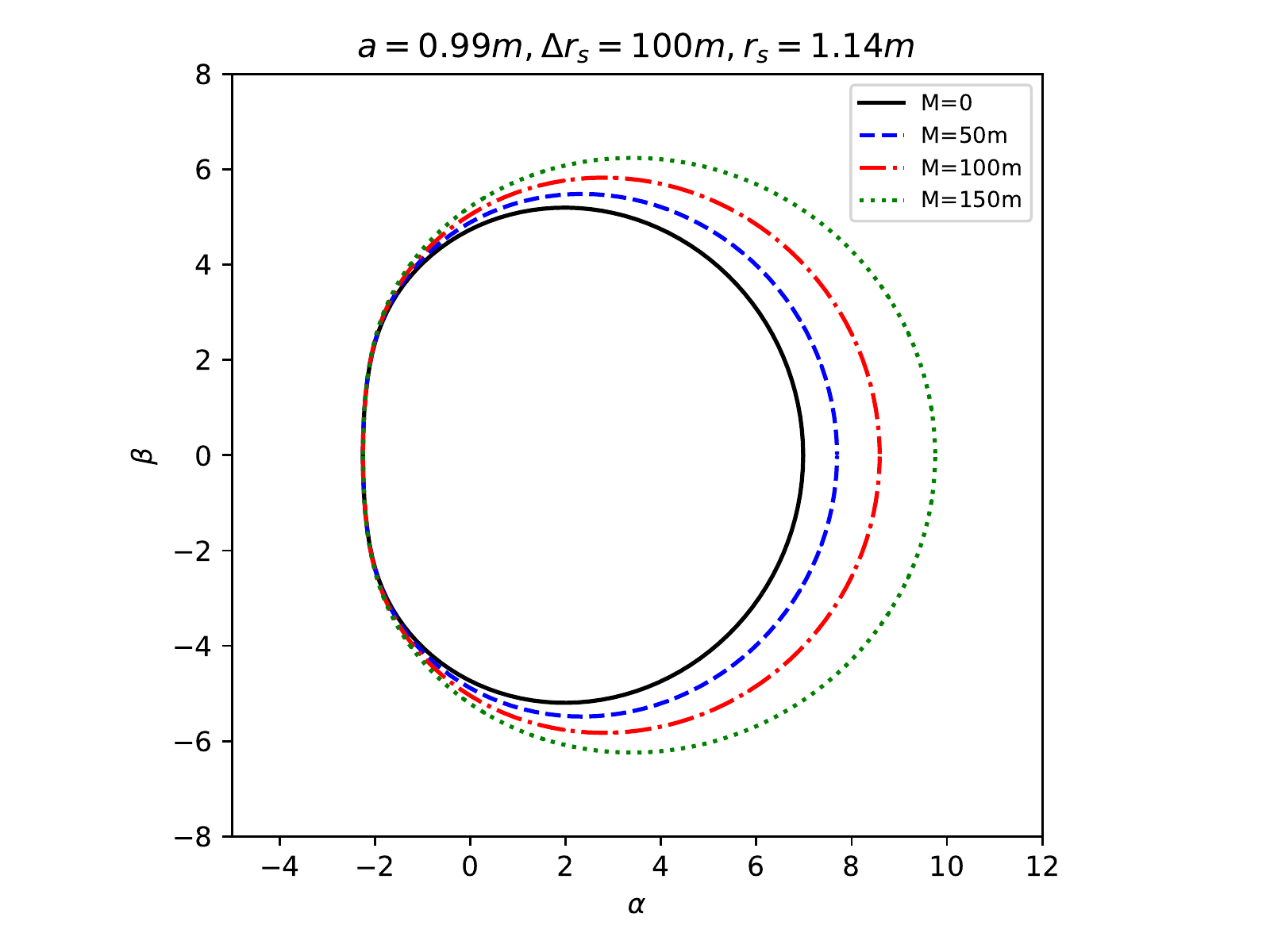}
    \end{minipage}
    \begin{minipage}{\columnwidth}
    \centering
    \includegraphics[width=\linewidth]{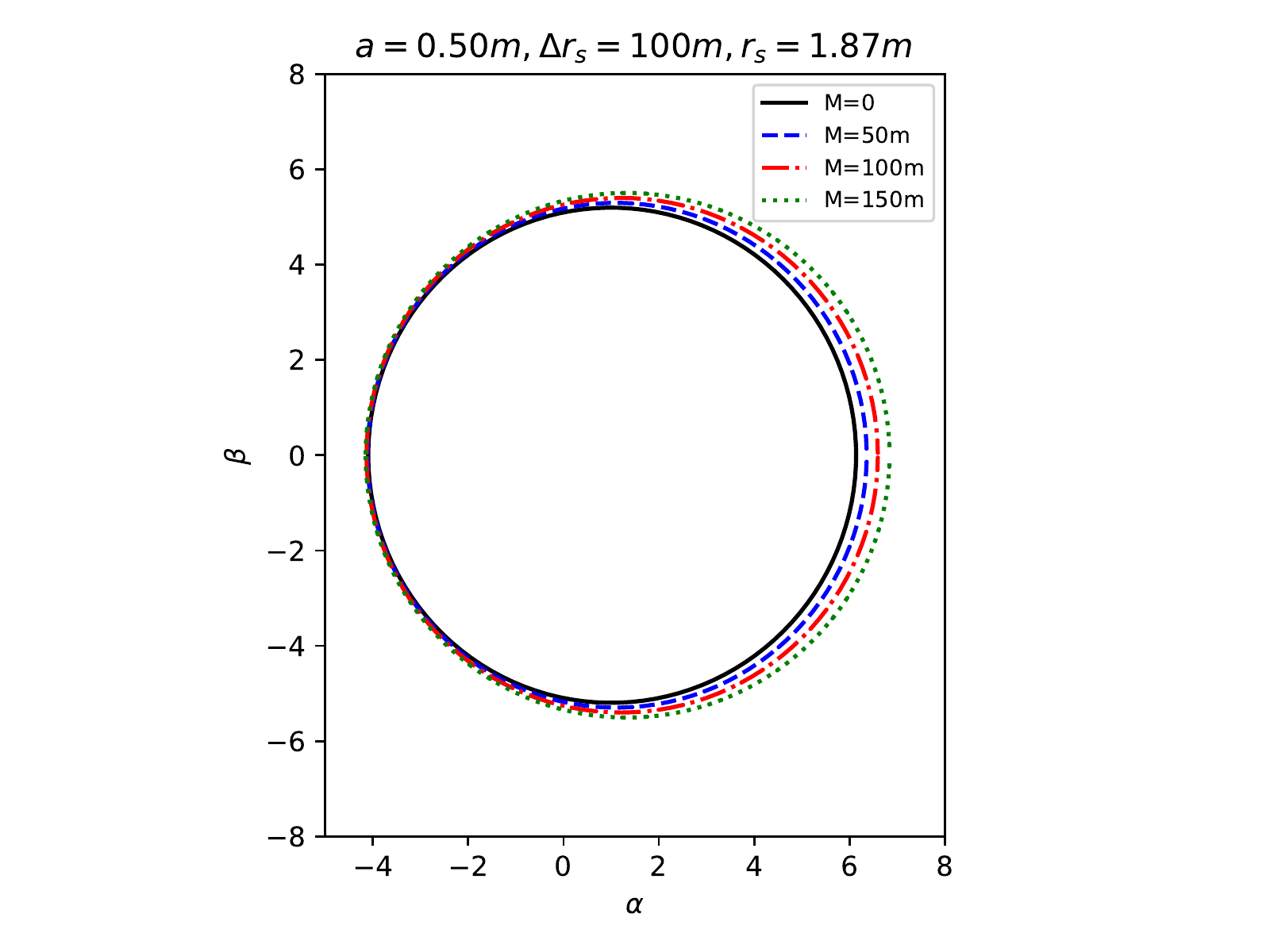}
    \end{minipage}
    \caption{Black hole shadow ($\theta_{o}=\pi/2$).}
    \label{fig7}
\end{figure}
This inconvenience is also true even with the approximation as $\Delta r_{s}\rightarrow\infty$, which reduces the above equation to the 4th power. By numerical calculations and satisfying Eq. \eqref{eq39n} we can locate the unstable photon orbits and obtain some insights as to what happens when $\Delta r_{s}\rightarrow\infty$ (i.e., low dark matter density). Unlike the time-like particles, Fig. \ref{fig6} reveals that high dark matter densitiy is needed in order to see deviations in the null orbits. In the near extremal case (top) the prograde is nearly unaffected, while the dark matter effect on the retrograde radius is to decrease its value relative to the Kerr case where $M=0$. For $a=0.50m$ (bottom) the change in the prograde orbit is evident. These changes, that the photon radius must decrease due to the presence of dark matter, agrees with the result in Ref. \cite{Konoplya2019}. As explained, the decrease in radius is due the dark matter's full effect (both under and above the photonsphere) causing a new orbital equilibrium.
\begin{figure}
    \begin{minipage}{\columnwidth}
    \centering
    \includegraphics[width=\linewidth]{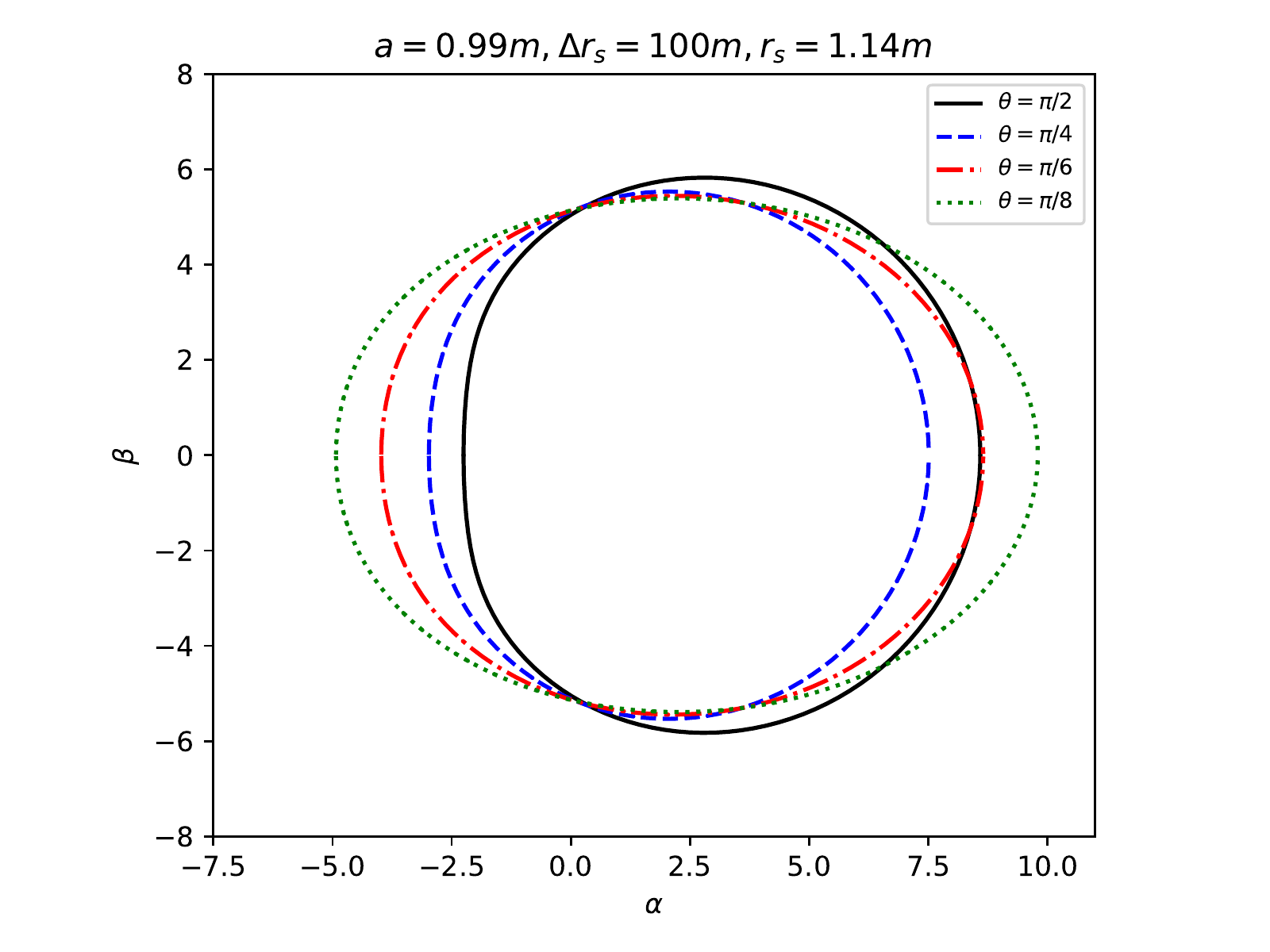}
    \end{minipage}
    \begin{minipage}{\columnwidth}
    \centering
    \includegraphics[width=\linewidth]{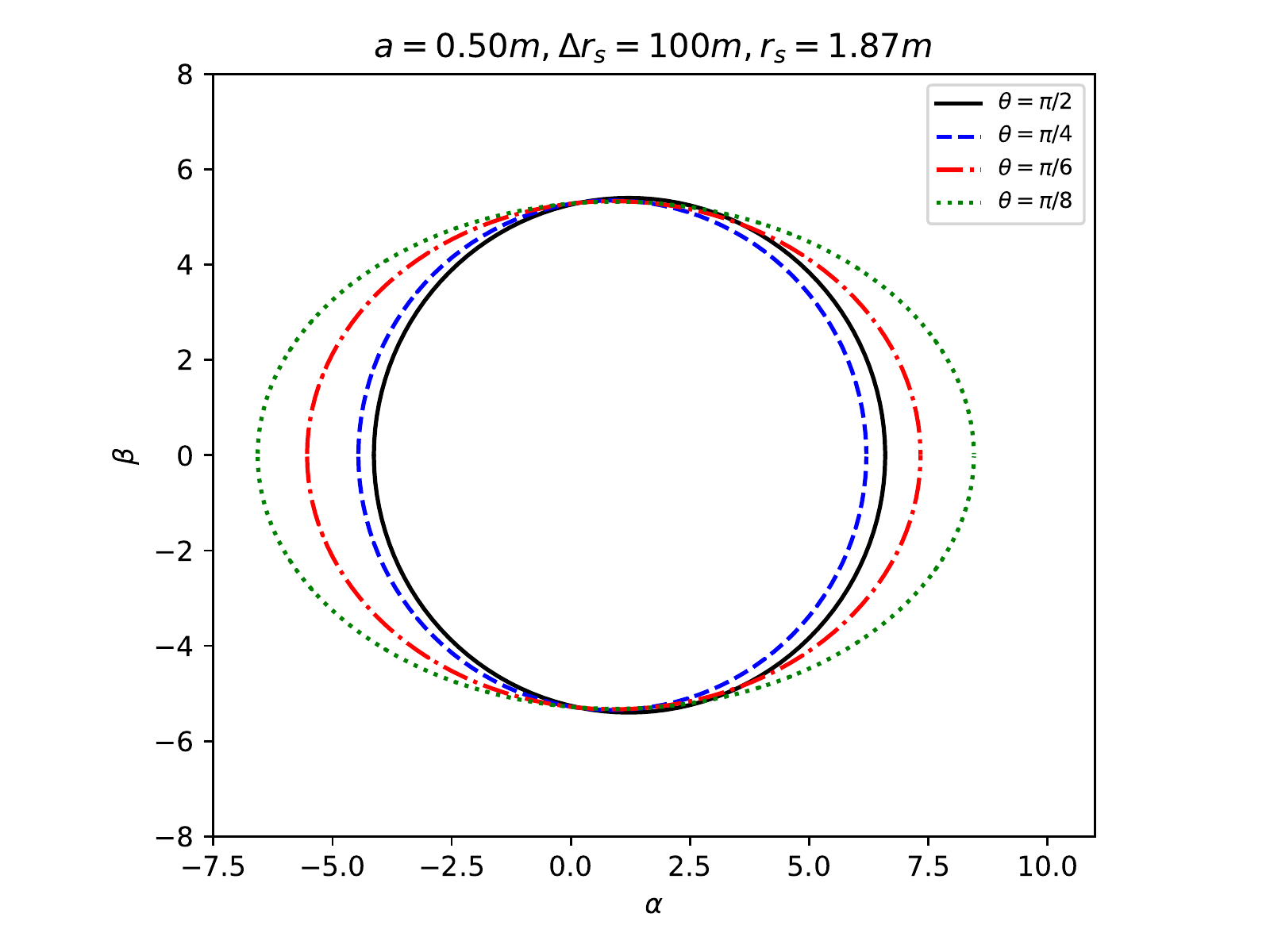}
    \end{minipage}
    \caption{Black hole shadow for different polar angle ($M=100m$).}
    \label{fig8}
\end{figure}

Any perturbations can lead the photons in the unstable orbit to escape the rotating black hole's gravitational influence. These photons will travel in the intervening space between the black hole and the remote observer. In our case, the photons will pass through the dark matter configuration. Hence, we expect a difference in the resulting shadow when vacuum, and space with dark matter, are compared. The method on deriving the celestial coordinates with respect to the Zero Angular Momentum Observers (ZAMO) is very well established. The celestial coordinates, in general, are given by \cite{johannsen2013photon}
\begin{align} \label{eq37}
&\alpha  =-r_{0}\frac{\xi }{
\zeta\sqrt{g_{\phi \phi }} \left( 1+\frac{g_{t\phi }}{g_{\phi \phi }}\xi
\right) }, \nonumber \\
&\beta  = r_{0}\frac{\pm 
\sqrt{\Theta (i)}}{\zeta\sqrt{g_{\theta \theta }} \left( 1+\frac{g_{t\phi }}{
g_{\phi \phi }}\xi \right) }
\end{align}
and in the limit $r\rightarrow\infty$, Eq. \eqref{eq37} reduces to
\begin{align} \label{eq38}
&\alpha=-\xi \csc \theta_0,   \nonumber \\
&\beta=\pm \sqrt{\eta +a^{2}\cos ^{2}\theta_0-\xi ^{2}\cot^{2}\theta_0}
\end{align}
where $\theta_o$ is the polar orientation of the remote observer with respect to the equatorial plane, while $\xi$ and $\eta$ are given by Eqs. \eqref{eq35} and \eqref{eq36}. Fig. \ref{fig7} shows how different dark matter density affects the black hole shadow. If there is no dark matter, we find the almost D-shaped contour of the Kerr black hole when the spin parameter is near extremal. When dark matter is present, the contour that represents the retrograde photon orbit bulges more as dark matter density increases. These contours perfectly agree with Fig. \ref{fig6}. In effect, this increases the radius of the shadow. The contour that represents the prograde orbit deviates less as dark matter density increases. With the given values of dark matter mass in the contour plot, it seems that the change in the size of the shadow is kind of exaggerated. It is only to demonstrate, however, how dark matter changes the size of the shadow. The D-shaped contour is not changed at all, hence, the fundamental properties of the rotating black hole is retained in the presence of dark matter. The bottom figure shows the shadow contour when the black hole spin is a bit lower.

When we consider different values of the polar angle $\theta_o$, Fig. \ref{fig8} shows how the remote observer sees the rotating black hole. As the observer gets near the poles, the shadow contour is becoming more of an ellipse-shaped.

\section{Shadow radius, radius distortion, and energy emission} \label{sec7}
\begin{figure}
    \centering
    \includegraphics[width=\linewidth]{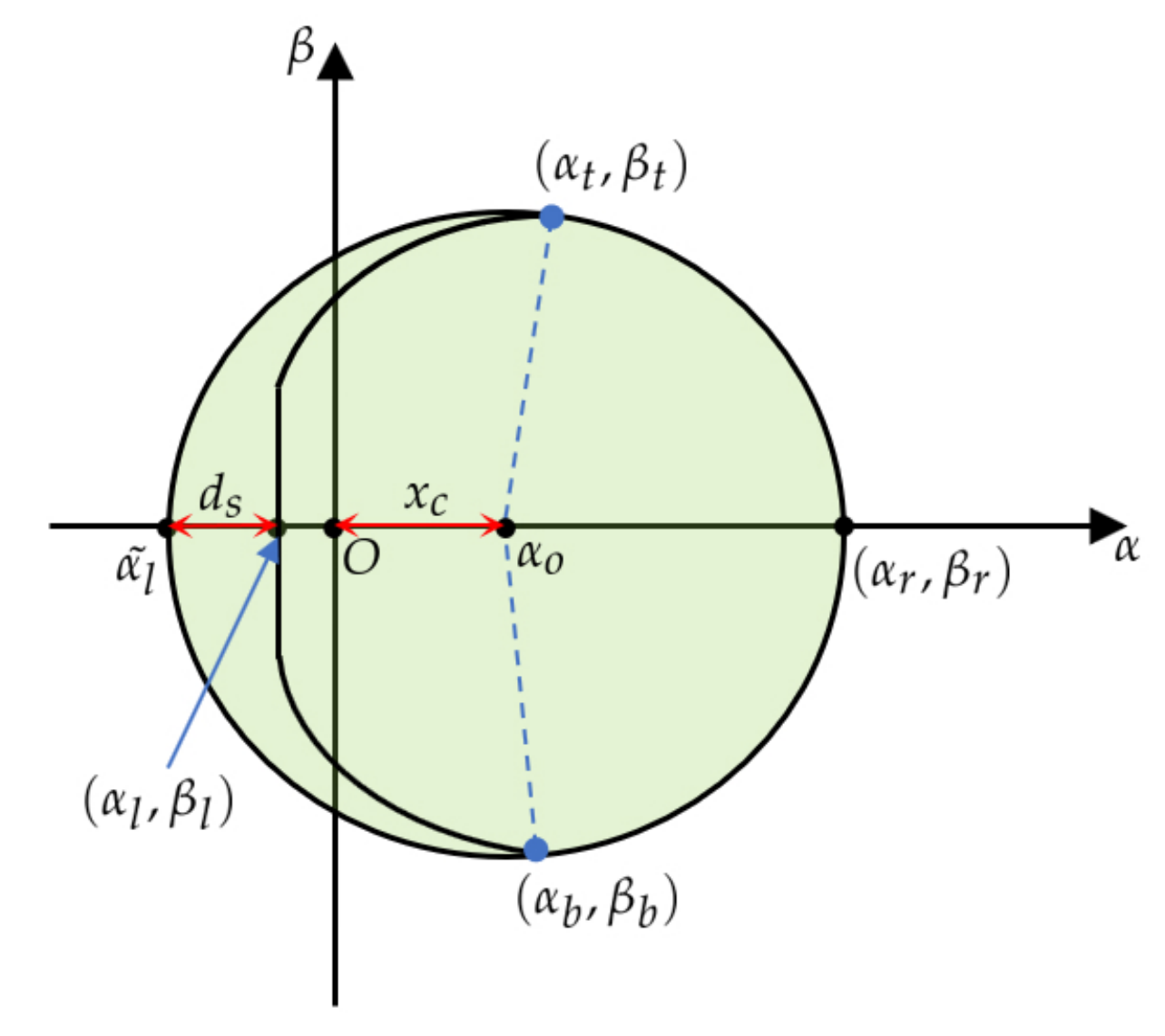}
    \caption{Schematic diagram of black hole shadow.}
    \label{fig9}
\end{figure}

The shadow radius and radius distortion parameter are very well known observables that are useful in extracting information about black hole shadows \cite{Kumar2018,Wei2019}. The schematic diagram of a black hole shadow overlapping a reference circle is shown in Fig. \ref{fig9} \cite{Hioki2009}. The radius of the shadow is then given by (for derivation, see Ref. \cite{Dymnikova2019})
\begin{equation} \label{eq39}
R_{s}=\frac{\beta_{t}^2+(\alpha_{t}-\alpha_{r})^2}{2|\alpha_{t}-\alpha_{r}|}
\end{equation}
where $\beta_{t}$, $\alpha_{t}$, and $\alpha_{r}$ can be found with the help of Eq. \eqref{eq38}.

\begin{figure}
    \begin{minipage}{\columnwidth}
    \centering
    \includegraphics[width=\linewidth]{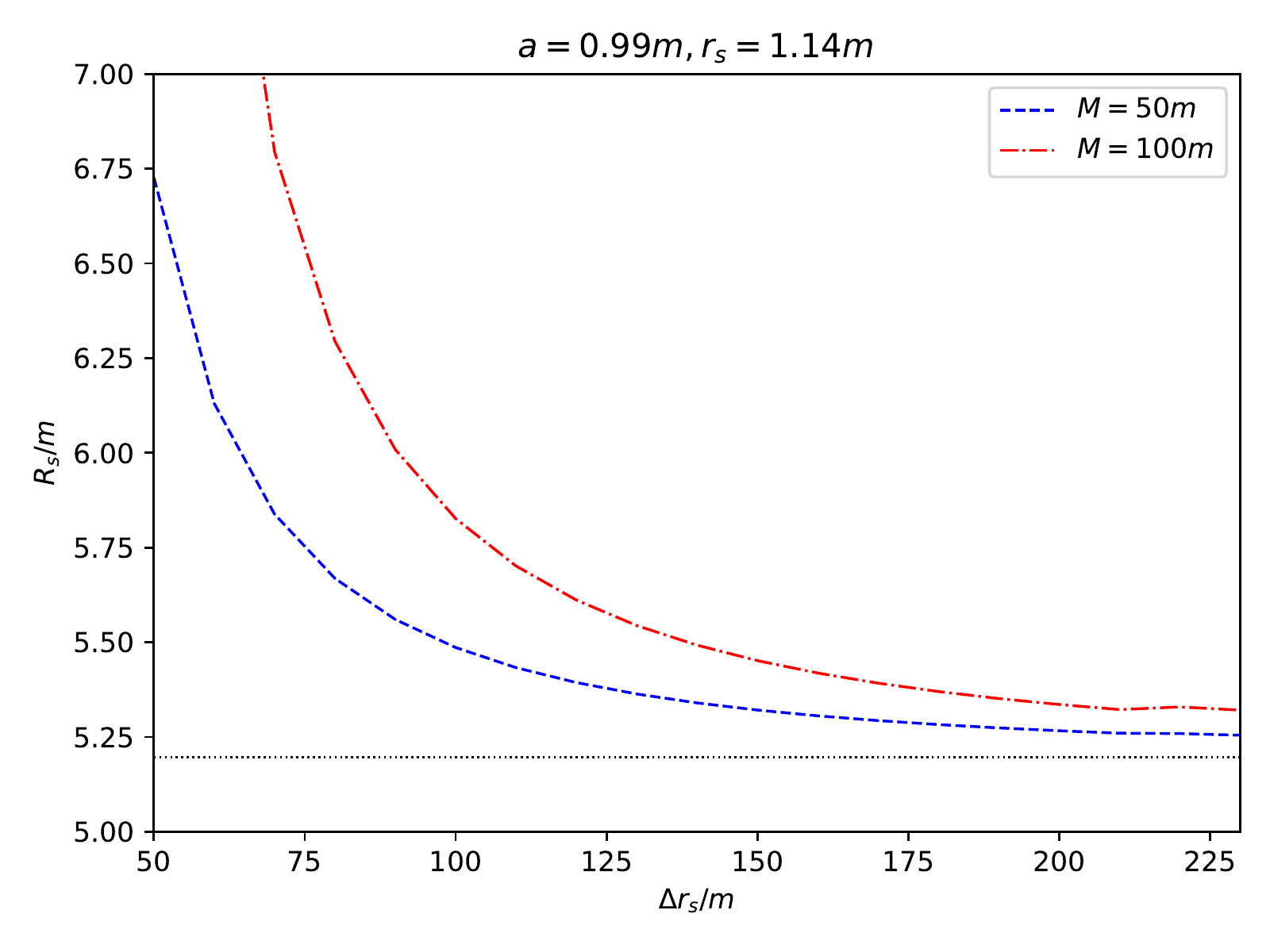}
    \end{minipage}
    \begin{minipage}{\columnwidth}
    \centering
    \includegraphics[width=\linewidth]{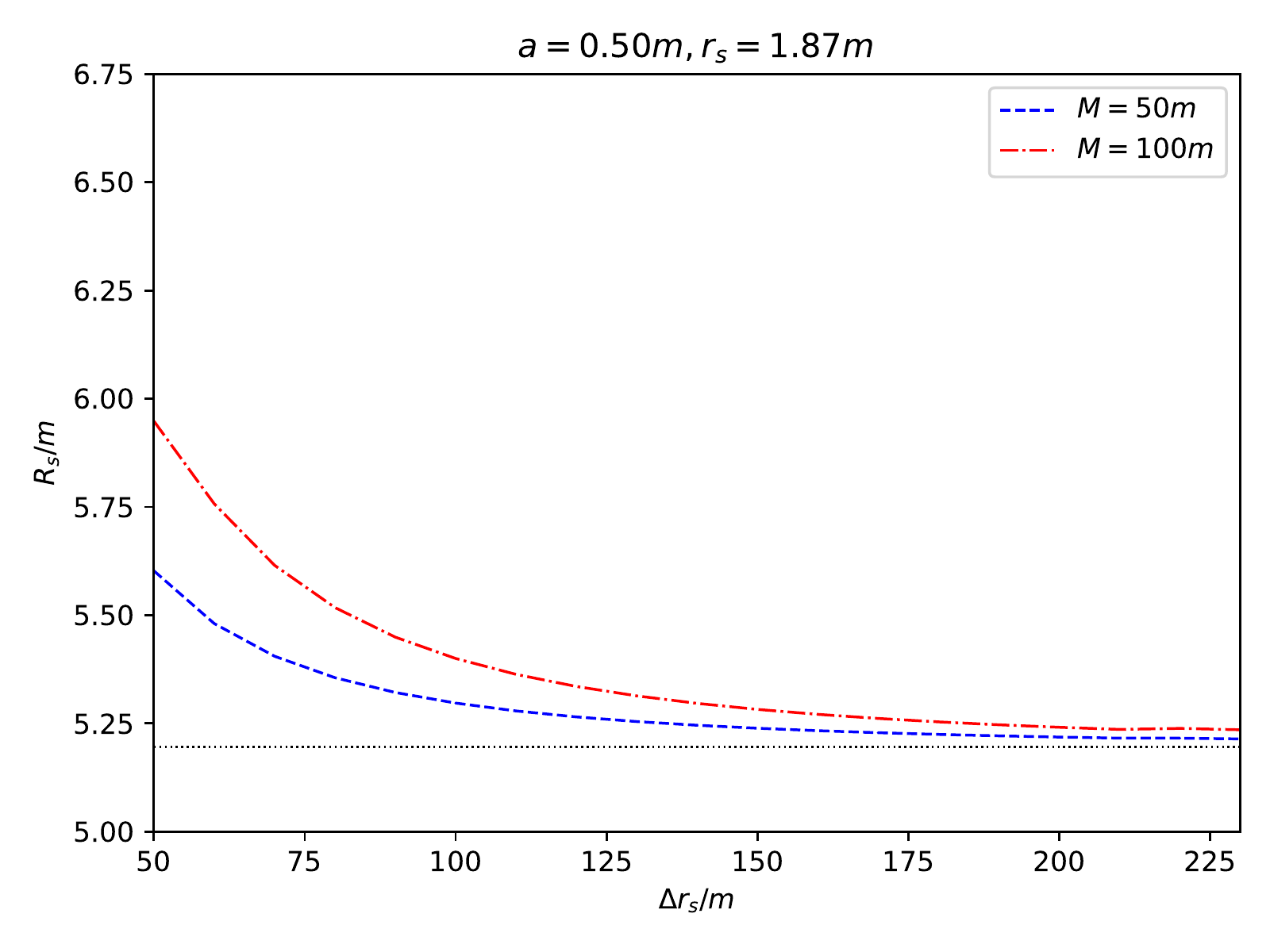}
    \end{minipage}
    \caption{Shadow radius.}
    \label{fig10}
\end{figure}

The shadow distortion is defined as $d_{s}=\tilde{\alpha}_{l}-\alpha_{l}$. The radius distortion parameter, in terms of shadow radius, can then be expressed as
\begin{equation} \label{eq40}
\delta_{s}=\frac{d_{s}}{R_{s}}=\frac{\tilde{\alpha}_{l}-\alpha_{l}}{R_{s}}
\end{equation}
Fig. \ref{fig10} shows how dark matter affects the shadow radius. The black dotted horizontal line represents the Schwarzschild case ($M=0$). Indeed, dark matter increases the shadow radius and such increase is also amplified by the black hole's spin parameter $a$. For both cases in the figure, the curve is asymptotic to the Schwarzschild case when $\Delta r_{s}\rightarrow\infty$. Due to Eq. \eqref{eqr^5} and the complexity looming in Eq. \eqref{eq39}, we emphasize again that it is inconvenient to derive a formula to estimate the effective radius of the dark matter halo in order to have considerable effect on the shadow radius. This is unlike the Schwarzschild scenario where the estimate $\Delta r_{s}=\sqrt{3mM}$ was easily attained because $r_{ph}$ can be derived analytically as well as the expression for the shadow radius.

\begin{figure}
    \begin{minipage}{\columnwidth}
    \centering
    \includegraphics[width=\linewidth]{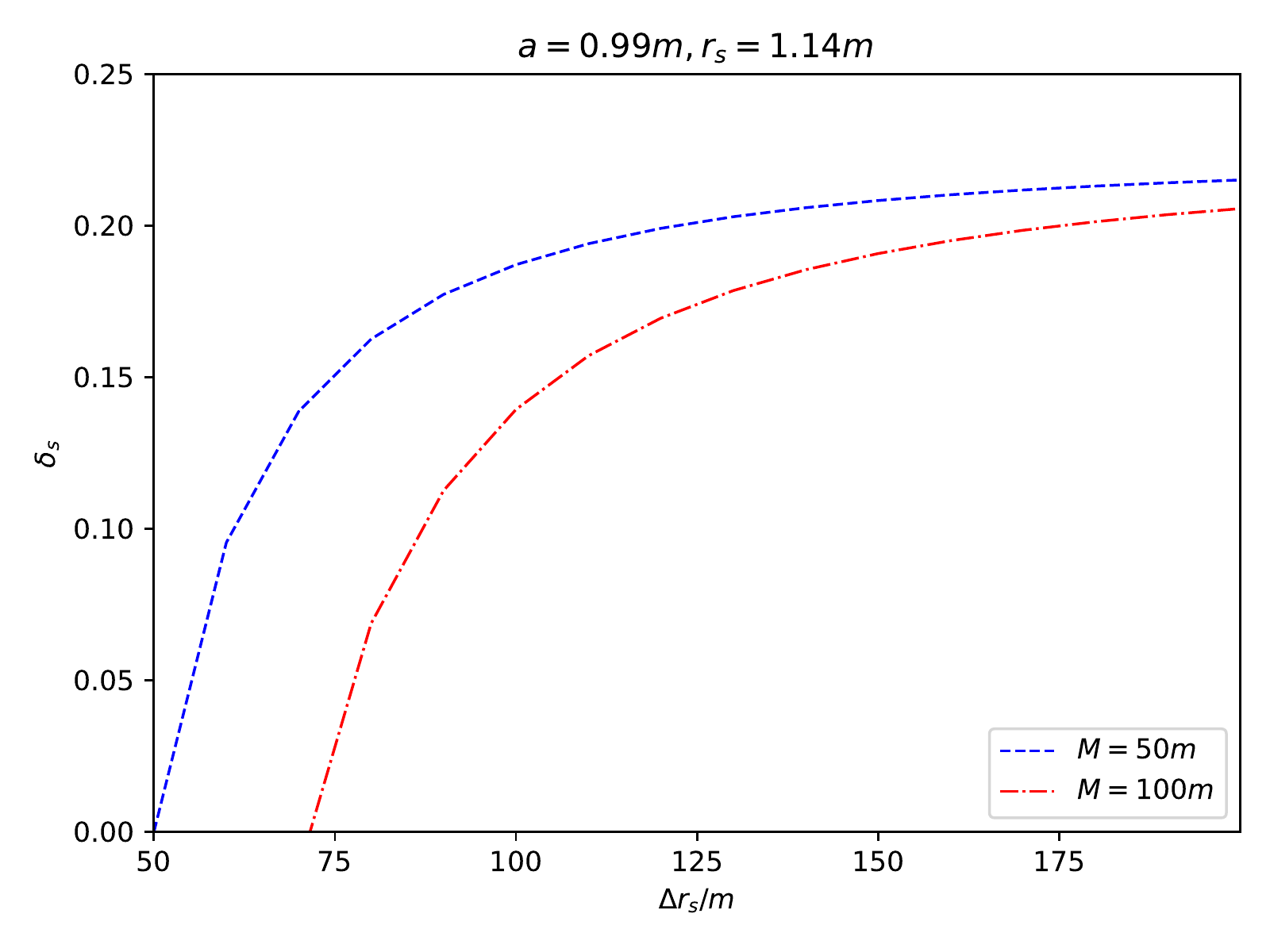}
    \end{minipage}
    \begin{minipage}{\columnwidth}
    \centering
    \includegraphics[width=\linewidth]{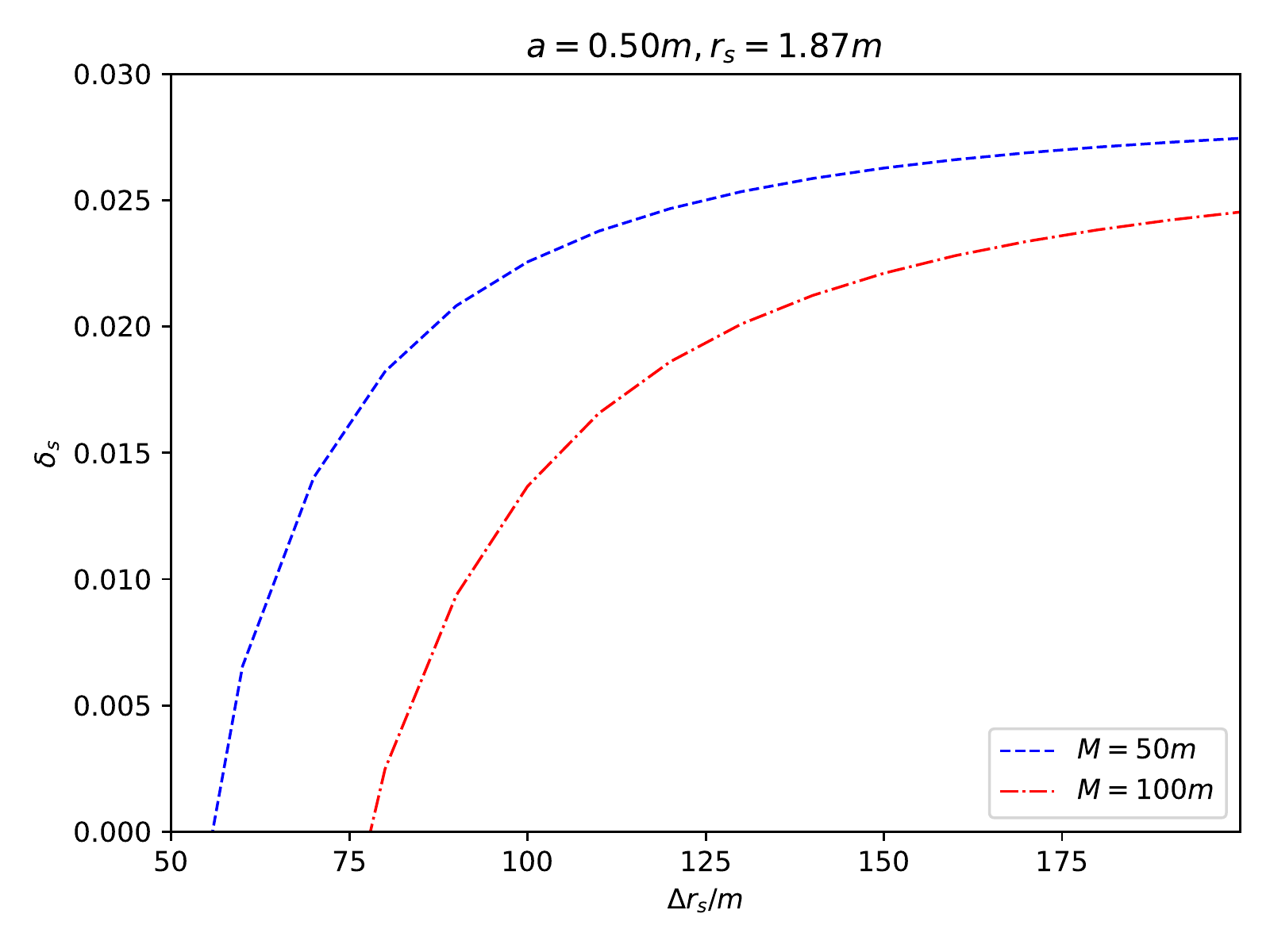}
    \end{minipage}
    \caption{Radius distortion parameter.}
    \label{fig11}
\end{figure}

The radius distortion parameter is plotted in Fig. \ref{fig11}. Here, we see the agreement in Fig. \ref{fig7} because as the spin parameter decreases, the more the shadow becomes close to a perfect circle. The radius distortion is indeed greater when the black hole spin is near the extremal case, which is also amplified by dark matter effect.

\begin{figure}
    \centering
    \includegraphics[width=\linewidth]{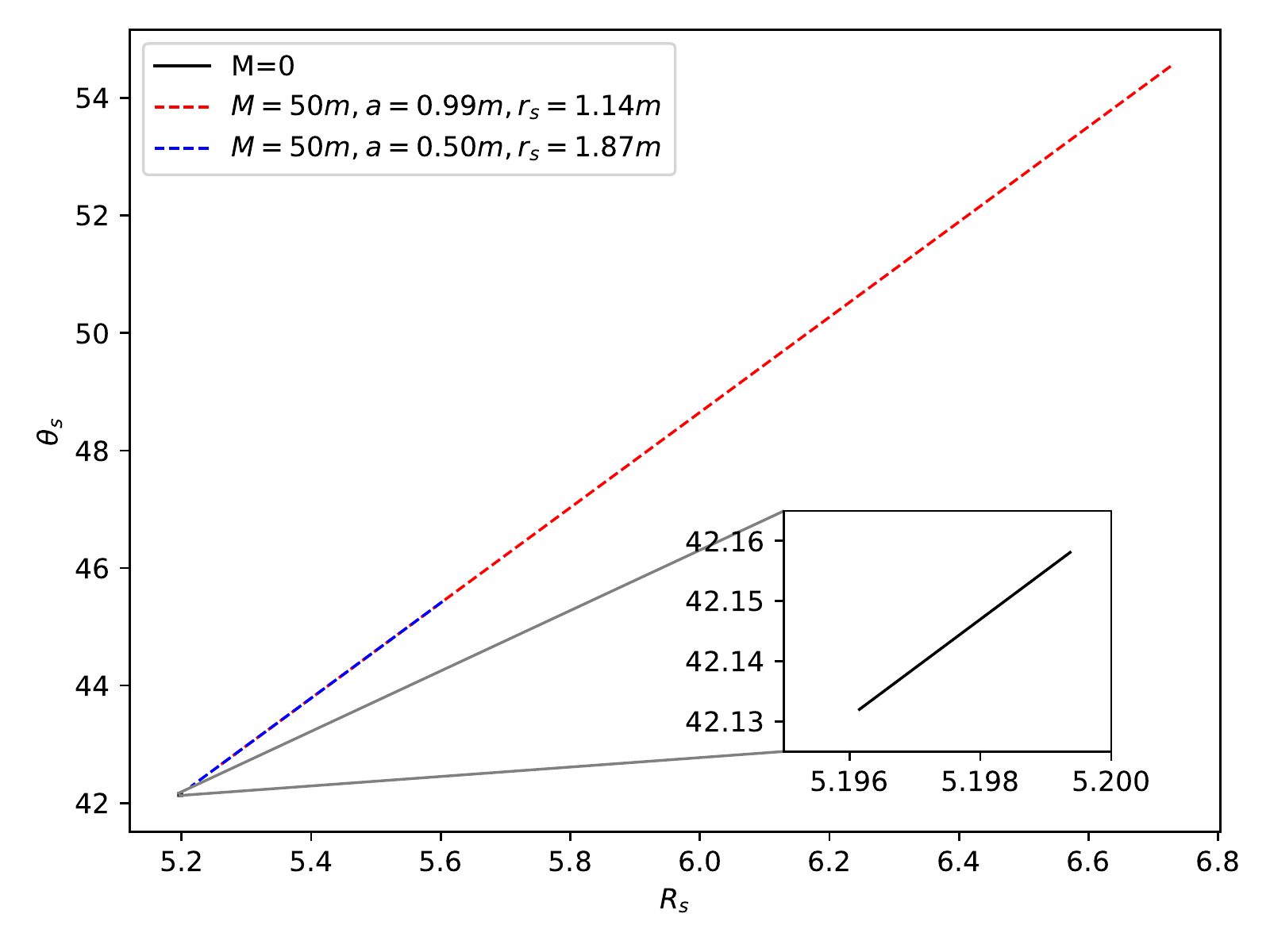}
    \caption{Angular diameter.}
    \label{fig12}
\end{figure}

We can also use the shadow radius to determine the angular diameter of the rotating black hole. The angular radius is given by
\begin{equation} \label{eq41}
\theta_{s}=9.87098\times10^{-3} \frac{R_{s}m}{D}
\end{equation}
where $m$ must be measured in terms of solar mass and $D$ in parsec. Let's consider supermassive black hole in M87 galaxy with mass $m=6.9\times10^{9}M_{\odot}$ and its distance from Earth is $D=16.8$Mpc. Fig. \ref{fig12} shows the plot with and without dark matter. In general, not only the dark matter influences the increase in angular diameter, but also the spin parameter. Even for $M=50m$ and $\Delta r_{s}=100m$, the angular diameter increases drastically as $a$ increases.
\begin{figure}
    \centering
    \includegraphics[width=\linewidth]{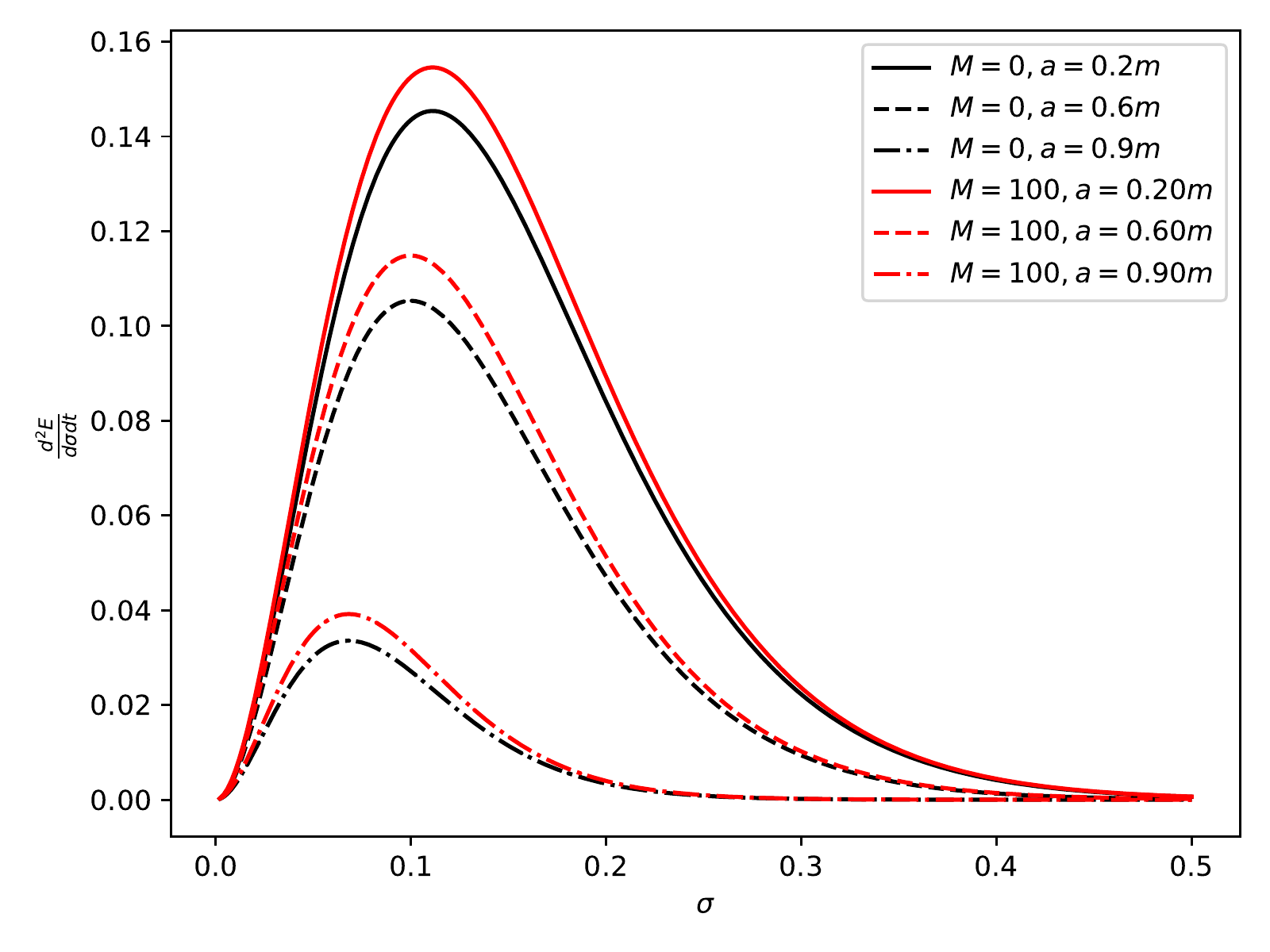}
    \caption{Energy Emission rate.}
    \label{fig13}
\end{figure}

The energy emission rate of a black hole is defined as
\begin{equation} \label{eq42}
\frac{d^{2}E}{d\sigma dt}=2\pi^{2}\frac{\Pi_{ilm}}{e^{\sigma/T}-1}\sigma^{3}
\end{equation}
where $T$ is the black hole temperature. Following Ref. \cite{Jusufi2019}, the temperature is given by
\begin{equation} \label{eq43}
T=\frac{r_{h}}{4\pi(r_{h}^{2}+a^{2})^{2}}\left[2a^{2}(f(r_{h})-1)+r_{h}(r_{h}^{2}+a^{2})f'(r_{h})\right]
\end{equation}
in which $r_h$ is the event horizon radius and $f(r_h)=-g_{tt}$ in the metric Eq. \eqref{eq20}. For a remote observer, the area of the shadow is approximately equal to the high energy absorption cross-section which oscillates around a constant, $\Pi_{ilm}=\pi R_{s}^2$. Fig. \ref{fig13} shows how the energy emission rate changes if the rotating black hole is surrounded by dark matter. As shown, the effect is evident when the dark matter mass is high ($M=100m$), where the peak frequency increases in the vertical axis. Hence, the effect of dark matter is to increase the energy emission rate near the event horizon. Dark matter also has a negligible effect on the photon's peak frequency because shifting to a lower or higher frequency is not so evident, even in the case of high dark matter density.

\section{Conclusion} \label{sec8}
In this paper, we extended the study in Ref. \cite{Konoplya2019} to a rotating case by utilizing the Newman-Janis algorithm. Focusing only on the interesting consequences of the second condition in Eq. \eqref{eq4}, and the case where $r_{s}=r_{h}$, we found that it requires high dark matter density to have considerable deviations in the horizons, ergosphere, as well as the null geodesics. Due to the complexity of how the shadow radius is defined in the Kerr case, it remains inconvenient to derive the necessary thickness $\Delta r_{s}$ for a notable change in the shadow radius to occur relative to an observer inside the shell. Hence, the result for $\Delta r_{s}$ in Refs. \cite{Konoplya2019,Pantig2020} remains a good estimate. New to this study is the analysis of the time-like orbits, and how it is affected by dark matter mass. We showed that time-like geodesics are very sensitive to dark matter effects because the location of the ISCO radius drastically changes even in a very low dark matter density environment. Other types of orbits are also seen to be affected by dark matter. The Penrose process is also shown to remain unaffected by dark matter.

Future research direction may include studying the effect of non-spherical dark matter distribution, or with a different expression for the density function. Further, one can also explore a more realistic model where the mutual influence between the black hole and dark matter is present.

\bibliography{references}

\end{document}